\documentclass[10pt, final]{IEEEtran}

\usepackage{lineno}
\usepackage{amsthm}
\usepackage{amsmath}
\usepackage[ruled]{algorithm2e}
\usepackage{mathrsfs}
\usepackage{cite}
\usepackage{bm,epsfig,amsthm,url}
\usepackage{indentfirst}
\usepackage{amssymb}
\usepackage{amsfonts}
\usepackage{epstopdf}
\usepackage[caption=false,font=footnotesize]{subfig}
\usepackage{xcolor}
\usepackage{mathtools}

\usepackage{hyperref}

\usepackage{microtype}

\theoremstyle{definition}

\newtheorem{theorem}{Theorem}

\newtheorem{remark}{Remark}
\newtheorem{assumption}{Assumption}

\usepackage{booktabs}
\usepackage{makecell}
\usepackage{graphicx}

\DeclareMathOperator*{\argmin}{arg\,min}

\begin{document}

\title{Performance-Oriented Design for Intelligent Reflecting Surface Assisted Federated Learning}

\author{Yapeng~Zhao,~\IEEEmembership{Graduate Student Member,~IEEE},
Qingqing~Wu,~\IEEEmembership{Senior Member,~IEEE},
Wen~Chen,~\IEEEmembership{Senior Member,~IEEE},
Celimuge~Wu,~\IEEEmembership{Senior Member,~IEEE}, and H.~Vincent~Poor,~\IEEEmembership{Life Fellow,~IEEE}
}

\maketitle

\vspace{-20pt}
\begin{abstract}
\textls[-1]{To efficiently exploit the massive amounts of raw data that are increasingly being generated in mobile edge networks, federated learning (FL) has emerged as a promising distributed learning technique by collaboratively training a shared learning model on edge devices. 
The number of resource blocks when using traditional orthogonal transmission strategies for FL linearly scales with the number of participating devices, which conflicts with the scarcity of communication resources. To tackle this issue, over-the-air computation (AirComp) has emerged recently which leverages the inherent superposition property of wireless channels to perform  \textit{one-shot} model aggregation. However, the aggregation accuracy in AirComp suffers from the unfavorable wireless propagation environment. In this paper, we consider the use of intelligent reflecting surfaces (IRSs) to mitigate this problem and improve FL performance with AirComp. Specifically, a novel performance-oriented long-term design scheme that integrated design multiple communication rounds to minimize the optimality gap of the loss function is proposed. We first analyze the convergence behavior of the FL procedure with the absence of channel fading and noise. Based on the obtained optimality gap which characterizes the impact of channel fading and noise in different communication rounds on the ultimate performance of FL, we propose both online and offline schemes to tackle the resulting design problem. Simulation results demonstrate that such a long-term design strategy can achieve higher test accuracy than the conventional isolated design approach in FL. Both the theoretical analysis and numerical results exhibit a ``later-is-better'' principle, which demonstrates the later rounds in the FL procedure are more sensitive to aggregation error, and hence more resources are required over time.}
\end{abstract}
\vspace{-3pt}
\vspace{-5pt}

\section{Introduction}
Recent years have witnessed a significant increase in artificial intelligence (AI) applications, such as image recognition and natural language processing. Moreover, some delay-sensitive applications of AI like autonomous driving, require timely processing of real-time data. Conventional approaches require a data center to collect all the raw data for centralized model training. However, collecting massive amounts of data from distributed devices incurs high latency and very high energy and bandwidth cost \cite{KaibinHuang_Mag20_EdgeAI, YuanmingShi_JSAC22_EdgeAI, YuanmingShi_Survey20_EdgeAI}. Furthermore, the computation resources of edge devices are wasted in such a centralized model, and potential privacy violations exist since local data is collected and processed at the central server. This has thus motivated the migration of AI applications from the center of networks to the network edge, where implicit knowledge can be locally distilled to provide timely and economical intelligence as well as improved privacy.  Chief among techniques for edge AI is federated learning (FL), in which local model parameters or gradients are exchanged instead of raw data \cite{FL_Survey20, MingzheChen_JSAC21_FL_survey}.

Since the training procedure in a state-of-the-art deep neural network (DNN) may involve millions of parameters \cite{FL_Survey20}, uplink communication overhead incurred during the iterative model update process becomes a critical bottleneck for FL given the limited communication bandwidth in practice. To address this issue, there has been considerable interest in communication-efficient uploading strategies for FL \cite{Poor_PNAS}. Conventionally, edge devices are orthogonally scheduled to upload their local model parameters/gradients, and the edge server sequentially decodes the received signals. However, the radio resources required in such orthogonal multiple access (OMA) protocols scale significantly with the number of participating devices and the model dimension. 
Therefore, different strategies have been proposed to alleviate this communication overhead, e.g., device sampling based on the network topology \cite{DeviceSampling}, model sparsification through the time correlations \cite{ISIT21_Sparsification}, and gradient quantization via lossy compression \cite{NIPS17_QSGD, PMLR18_SIGNSGD}. However, all of these schemes sacrifice performance in order to deal with the limited communication resources. 
As an alternative, the recently aroused over-the-air computation (AirComp) technique has shown its efficiency in the \textit{one-shot} aggregation of simultaneously transmitted local model parameters/gradients as compared to OMA protocols \cite{ AirComp_survey}. AirComp essentially turns the air into a computer by leveraging the inherent waveform superposition property of wireless channels. Particularly, it allows all the participating devices in FL to access all the radio resources simultaneously instead of only a fraction of them as in conventional OMA schemes. 
Moreover, its capability for \textit{one-shot} function computation has been validated in a series of works, such as the foundational study from an information theoretic perspective \cite{Nazer_TIT07, Nazer_TIT11}, performance analysis for various systems from the signal processing perspective \cite{KaibinHuang_IoT19_MIMO_AirComp, MinFu_UAV_AirComp}, and prototype validation in practical implementation \cite{AirComp_Implement21}. 

AirComp empowered FL systems have been comprehensively investigated recently, including how to accelerate the convergence of FL under tight resources constraints (e.g., communication and computation resources) \cite{MingzheChen_TWC21_ConvergenceTime} or reduce resources consumption while maintaining satisfactory learning performance \cite{ZhaohuiYang_TWC21_FL}. Specifically, there have been two main lines of research in existing works: one is system optimization from the communication perspective including device scheduling \cite{YuanmingShi_TWC21_FL_IRS}, transceiver design \cite{ XiaowenCao_JSAC22}, and parameter/gradient compressing \cite{KaibinHuang_TWC21_OneBit_FL}, the other is the inherent design of hyperparameters in the learning procedure, such as the learning rate \cite{ZhaohuiYang_JSAC21_Learning_rate} and batch size \cite{JSAC21_Batchsize_FL}. For example, the authors in \cite{YuanmingShi_TWC21_FL_IRS} proposed a two-step framework for joint device scheduling and receive beamformer design to improve the test accuracy. The local learning rate at edge devices was optimized in \cite{ZhaohuiYang_JSAC21_Learning_rate} to combat the distortion induced by fading channels. Although AirComp has been envisioned and further validated to be a promising scalable model aggregation solution in FL, a crucial defect obstructs its implementation in practice. Specifically, the devices whose channels are deeply faded dominate the aggregation error in AirComp-based FL \cite{KaibinHuang_IoT19_MIMO_AirComp, MinFu_UAV_AirComp}. This is because the \textit{one-shot} aggregation demands that the signals transmitted by different devices be equally superimposed at the receiver, and thus the remaining devices with better channel conditions have to reduce their transmit power in order to perform signal alignment at the receiver. To this end, the strength of the obtained signal is diminished, and the aggregation accuracy is more sensitive to the inherent noise. 
Therefore, the unfavorable propagation environment inevitably limits the performance of AirComp-based FL. Fortunately, intelligent reflecting surfaces (IRSs), a promising technology for the beyond fifth-generation (B5G) and the future sixth-generation (6G) network, has shown its potential to overcome this detrimental effect \cite{WQQ_IRS_Survey, rajatheva2020scoring, MengHua}. By intelligently tuning signal reflections via a large number of low-cost passive reflecting elements, IRSs are capable of dynamically altering wireless channels to enable precise model aggregation, thus enhancing the FL performance. 
Recent works have demonstrated the effectiveness of IRSs in improving the performance of AirComp-based FL, e.g., weak channels were enhanced by IRSs to involve more devices in collaborative model training while sustaining precise aggregation thus accelerating the model convergence in \cite{YuanmingShi_TWC21_FL_IRS, HangLiu_TWC21_FL_IRS, FL_Multi_IRS}. Besides, IRS was utilized to flexibly adjust the decoding order of heterogeneous data to serve both the FL and non-orthogonal multiple access (NOMA) users on the same time-frequency resource in \cite{WanliNi_TWC22_FL_NOMA} and minimize the energy consumption of engaged devices in \cite{ICASSP21}.

However, the communication system design in the existing literature has not been tailored to the inherent characteristics of FL. The existing system design schemes in AirComp-based FL mainly focused on the isolated system design in each communication round \cite{YuanmingShi_TWC21_FL_IRS, FL_Multi_IRS}.
In fact, FL is a long-term process consisting of many progressive learning rounds that collaboratively determine the ultimate learning performance \cite{OptimalityGap_TWC21, CongShen_Mag21}. Different learning rounds may have varying significance toward the convergence rate and the final model accuracy due to this intrinsic nature. Hence, resources need to be balanced among different iterations in FL by analyzing the collective impact of successive communication rounds on the ultimate performance.
The isolated resource allocation strategy in the existing works equally treats each learning round, which inevitably results in performance loss.

In this paper, we consider an AirComp-based FL system, in which an IRS is employed to configure a favorable wireless channel to provide precise model aggregation in each iteration. 
For the first time, we propose a performance-oriented long-term design approach to fully unleash the available communication resources, thus obtaining higher test accuracy. 
The first step toward resource allocation across different learning rounds is to evaluate the impact of resources on the overall learning accuracy and convergence. To this end, we characterize the optimality gap of the loss function in arbitrary communication rounds, which unveils the relationship between communication accuracy in each communication round and ultimate learning performance. 
Accordingly, a long-term system design scheme is proposed to minimize the optimality gap via integrated design over multiple communication rounds. Different from existing works that focused on the isolated design in each communication round, it is interesting to find that such an integrated approach exhibits a \textit{later-is-better} principle, which confirms that the later rounds in the FL procedure are more sensitive to the aggregation error, and hence more resources are required over time. 
The main contributions of this paper can be summarized as follows.
\begin{itemize}
	\item We derive the general optimality gap of the loss function, which characterizes the impact of gradient aggregation errors in different communication rounds on the convergence performance of AirComp-based FL. From the obtained optimality gap, it is observed that within a finite number of communication rounds, the aggregation errors in later rounds have a greater impact on the optimality gap than those in earlier rounds. Hence, the later rounds are more essential to the learning performance of FL, and more resources need to be allocated to them. This observation provides important guidance for practical system design. However, the conventional isolated system design approach neglects the significant property and treats all the communication rounds equally.
	\item To enhance the ultimate performance of FL, we integrated design multiple communication rounds to minimize the optimality gap and establish the long-term system design schemes. We first propose an offline design approach in which the entire FL procedure is sequentially decomposed based on the available lookahead information, and then leverage the Lyapunov framework \cite{Lyapunov_Book} to construct an online design problem over each communication round without foreseeing the future. The transceiver design and the IRS phase shift tuning are decoupled through the block coordinate descent (BCD) method. Furthermore, we propose an element-wise successive refinement algorithm with low complexity to guide the practical implementation of an IRS with discrete phase shift constraint.
	\item We conduct extensive simulations to evaluate the performance of the proposed performance-oriented design framework on the MNIST dataset \cite{MNIST} and the CIFAR-10 dataset \cite{Krizhevsky09} with convolution neural networks (CNNs). Simulation results show that the proposed long-term design schemes can achieve higher test accuracy than the conventional isolated design approach eventually, although it may lag behind in the beginning. Furthermore, it is observed that the online system design approach can acquire satisfactory performance compared to the offline solution without foreseeing the future, which further validates the \textit{later-is-better} principle. Simulation results demonstrate that the proposed online scheme is a practical and promising approach for FL system design.
\end{itemize}

The remainder of this paper is organized as follows. In Section II, we describe the FL model and the IRS-assisted AirComp framework. In Section III, we analyze the FL performance and accordingly formulate the performance-oriented optimization problem. In Section IV, we harness an efficient BCD approach to alternatively optimize transceiver and IRS phase shifts. In Section V, we present extensive numerical results to evaluate the proposed algorithm. Finally, this paper concludes in Section VI.

\emph{Notation:} Scalars are denoted by italic letters, and vectors and matrices are denoted by bold-face lower-case and uppercase letters, respectively. $\mathbb{R}^{m\times n}$ and $\mathbb{C}^{m\times n}$ denote the space of $m\times n$ real-valued and complex-valued matrices, respectively. For a complex-valued vector $\bm{x}$, ${\left\| \bm{x} \right\|}$ represents the  Euclidean norm of $\bm{x}$, ${\rm arg}(\bm{x})$ denotes the phase of $\bm{x}$, and ${\rm diag}(\bm{x}) $ denotes a diagonal matrix whose main diagonal elements are extracted from the vector $\bm{x}$. For a square matrix $\bm{S}$, $\mathrm{tr}(\bm{S})$ and $\bm{S}^{\mathrm{-1}}$ denote its trace and inverse, respectively, while ${\bm{S}} \succeq {\bm{0}}$ means that $\bm{S}$ is positive semi-definite, where $\bm{0}$ is a zero matrix of proper size. For any general matrix $\bm{A}$, $\bm{A}^{\mathrm{H}}$, $\mathrm{rank}(\bm{A})$, and $\bm{A}_{i,j}$ denote its conjugate transpose, rank, and $(i, j)$th entry, respectively. $\bm{I}_M$ denotes an identity matrix of size $M\times M$. $\jmath$ denotes the imaginary unit, i.e., $\jmath^2= -1$. $\mathbb{E}[\cdot] $ denotes the statistical expectation. $|\cdot| $ denotes the cardinality of a given set. ${\cal O}\left(  \cdot  \right)$ is the big-O computational complexity notation.

\section{System Model}
\begin{figure}[h]
\centering
\includegraphics[width=3in]{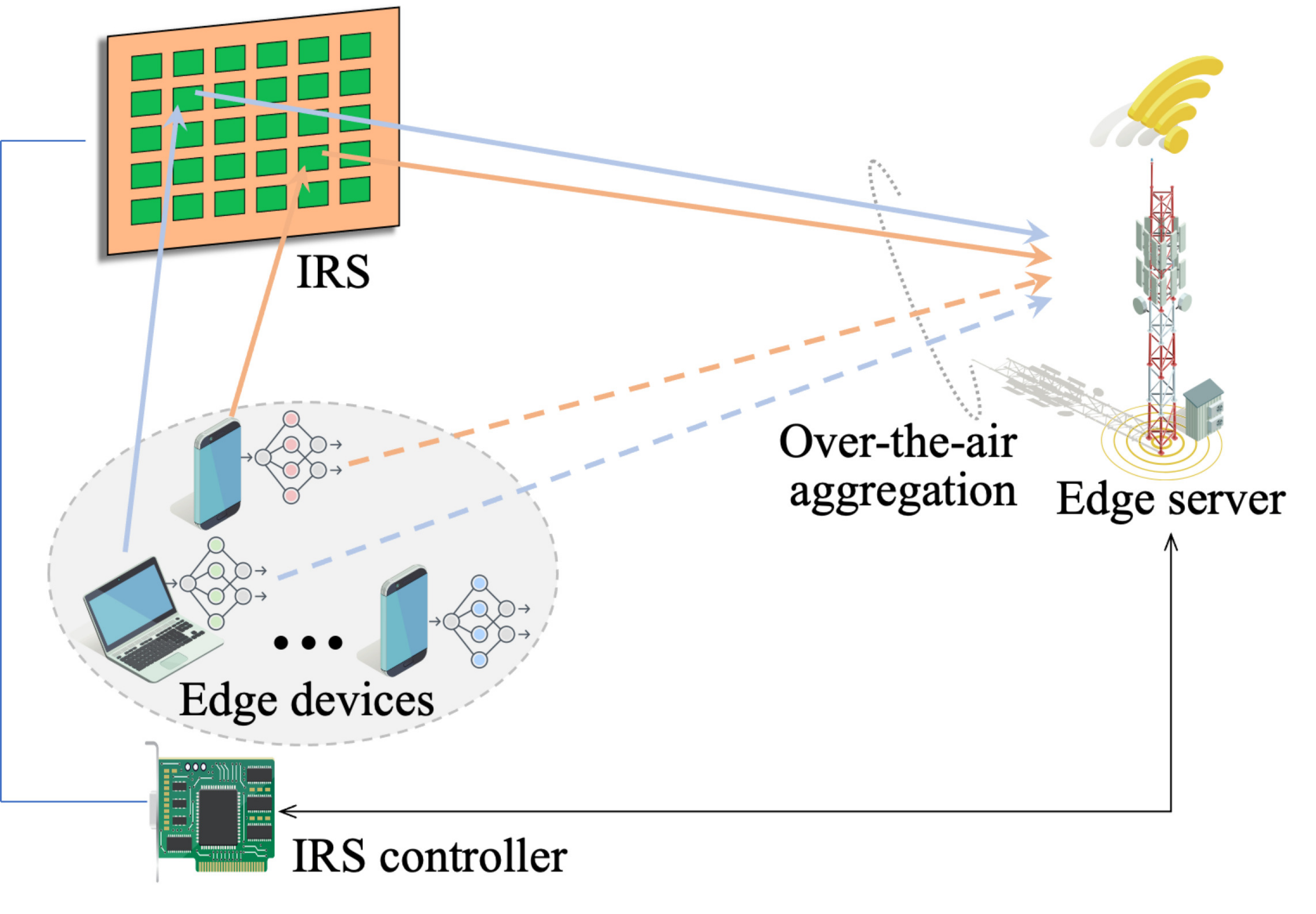}
\vspace{-10pt}
\caption{An IRS-assisted AirComp-based FL system.}
\label{F_FL}
\vspace{-10pt}
\end{figure}

As depicted in Fig. \ref{F_FL}, we consider an IRS-assisted FL system comprising of $K$ single antenna devices, an IRS with $N$ passive elements, and a base station (BS) with $M$ antennas serving as an edge server. Devices and the IRS are coordinated by the BS to realize the collaborative model training through periodical communication and computation. The details of the FL procedure and the IRS-assisted over-the-air aggregation are introduced as follows.

\vspace{-8pt}
\subsection{Federated Learning Procedure}

The goal of the FL system is to collectively acquire the desired parameter $\mathbf{w}^* \in \mathbb{R}^d $ that minimizes the global loss function with respect to the entire dataset. Instead of uploading all the raw data to the server, each device processes its local data in parallel and uploads the correspondingly generated local parameters or gradients to the server. We adopt the gradient sharing strategy in the FL procedure, i.e., the edge devices compute their local gradients and upload them to the edge server, and then the edge server broadcasts the global gradient which is computed based on the aggregated gradients to the edge devices for synchronizing local models' update.

Specifically, we assume that each edge device holds a subset of training data with $|\mathcal{D}_k|$ samples that are sampled i.i.d. from a source distribution $\mathcal{D}= \bigcup_{k\in \mathcal{K}} \mathcal{D}_k$, which is denoted by $\mathcal{D}_k = \{ \mathbf{u}_1, \mathbf{u}_2, \ldots, \mathbf{u}_{|\mathcal{D}_k|} \}$. Notation $\mathbf{u}_i$ represents a data sample, which can be represented as a set of feature vectors and its label in supervised learning. The regularized local loss function of device $k$ is given by 
\vspace{-5pt}	
\begin{equation}
	F_k(\mathbf{w}) = \frac{1}{|\mathcal{D}_k|} \sum_{\mathbf{u}_i \in \mathcal{D}_k} f(\mathbf{w}; \mathbf{u}_i) + \lambda'R(\mathbf{w}),
\end{equation}
where $R(\mathbf{w})$ is a strongly convex regularization function, $\lambda' \geq 0$ is a scaling hyperparameter. The global loss function evaluated at model parameter $\mathbf{w}$ is
\vspace{-5pt}
\begin{align}
	F(\mathbf{w}) = \frac{1}{D_{\textrm{tot}}}  \sum_{k=1}^K |\mathcal{D}_k| F_k(\mathbf{w}),
\end{align}
where $D_{\textrm{tot}} = \sum_k |\mathcal{D}_k|$.
This amounts to the regularized empirical average of the sample-wise loss functions on the global data set $\mathcal{D}$. The FL procedure aims to obtain the optimal $\mathbf{w}^*$ that minimizes the global loss function: 
\vspace{-3pt}
\begin{equation}
	\mathbf{w}^* = \argmin \, F(\mathbf{w}).
\end{equation}

To perform cooperative training among edge devices, the edge devices compute their local gradients by minimizing $F_k(\mathbf{w})$ in parallel, and then the BS obtains the global gradient based on gathered local gradients. Specifically, based on a randomly sampled mini-batch $\hat{\mathcal{D}}_k$ from the local dataset, each device $k$ computes a gradient vector $\mathbf{g}_{k}^{(t)}$ at communication round $t$ as
\begin{equation}
	\mathbf{g}_{k}^{(t)} \! \! = \! \! \nabla F_k(\mathbf{w}^{(t)}) \! = \! \frac{1}{|\hat{\mathcal{D}}_k|} \! \! \sum_{\mathbf{u}, v\in \hat{\mathcal{D}}_k} \! \! \! \nabla f(\mathbf{w}; \bm{u}, v) + \! \lambda \nabla R(\mathbf{w}^{(t)}),
\end{equation}
where $\mathbf{w}^{(t)}$ is the model parameter before updating. In this paper, we assume that all the mini-batch sizes are equal among devices.\footnote{The utilized FL framework can be readily extended to the case that devices are equipped with different data sizes. In this case, \eqref{e_global} can be revised to a weighted-average form.} Hence, the aggregated global gradient is
\begin{equation} \label{e_global}
	\mathbf{g}^{(t)} = \frac{1}{K} \sum_{k=1}^K  \mathbf{g}_{k}^{(t)}.
\end{equation}
Then the BS broadcasts the global gradient to edge devices for synchronously local model updating. We assume that downlink communication is ideal, hence each device can obtain precise global gradient $\mathbf{g}^{(t)}$ for local model updating as 
\begin{equation}
\mathbf{w}^{(t+1)}=\mathbf{w}^{(t)}-\alpha^{(t)} \cdot \overline{\mathbf{g}}^{(t)},
\end{equation}
in which $\alpha^{(t)}$ denotes the learning rate in the communication round $t$.

\subsection{Over-the-air Aggregation}

We adopt the AirComp technique to perform the communication-efficient gradient aggregation, and the IRS is deployed to assist the signal transmission from devices to the BS. 
Generally, the target function for aggregating the local gradient at the BS can be expressed as
\begin{align} \label{aggregating}
	\mathbf{g}^{(t)} = \chi^{(t)}(\varrho_k^{(t)}(\mathbf{g}_k^{(t)})),
\end{align}
where $\varrho_k^{(t)}(\cdot)$ denotes the pre-process function at device $k$ for normalization in round $t$, and $\chi^{(t)}(\cdot)$ denotes the post-process function for de-normalization at the BS. Hence, the transmit sequence is $\mathbf{s}_{k}^{(t)} = \varrho_k^{(t)}(\mathbf{g}_k^{(t)})$.
The baseband equivalent channels in communication round $t$ from the IRS to BS, from the device $k$ to IRS, and from the device $k$ to BS are denoted by $\bm{G}^{(t)} \in \mathbb{C}^{M \times N}$, $\bm{h}_{r,k}^{(t)} \in \mathbb{C}^{N \times 1}$ and $\bm{h}_{d,k}^{(t)} \in \mathbb{C}^{M \times 1}$, respectively. The CSI in current communication round is assumed to be perfectly known based on the various channel acquisition methods discussed in \cite{WQQ_IRS_Survey}. Due to the severe path loss, the power of the signals that are reflected by the IRS two or more times is assumed to be negligible \cite{WQQ_TWC19_IRS}. 
With the assistance of IRS, the signal received at the BS is given by
\begin{align}
	\bm{Y}^{(t)}&= \!\!\sum_{k \in \mathcal{K}^{(t)}}\left(\bm{h}_{d, k}^{(t)}+\bm{G}^{(t)} \bm{\Theta}^{(t)} \bm{h}_{r, k}^{(t)}\right) b_{k}^{(t)} (\mathbf{s}_{k}^{(t)})^{\mathrm{T}} \! + \!\boldsymbol{Z}^{(t)},
\end{align}
where $b_{k}^{(t)}\in \mathbb{C}$ is the transmit factor of device $k$ that controls the power consumption in round $t$, and $\bm{Z}^{(t)} \in \mathbb{C}^{M\times d} $ with each entry $z_i \sim \mathcal{CN}(0,\sigma_z^2 ) $ denotes the additional white Gaussian noise at the receiver. Besides, $\bm{\Theta}^{(t)}= \operatorname{diag}(\beta_1^{(t)} e^{\jmath\theta_1^{(t)}}, \ldots, \beta_N^{(t)} e^{\jmath\theta_N^{(t)}} ) $ denotes the diagonal phase shift matrix of the IRS, where $\theta_n^{(t)} \in [0,2\pi), \beta_n^{(t)} \in [0,1], \forall n \in \mathcal{N}, \forall t \in \mathcal{T}$ denote the phase shift and the amplitude reflection coefficient on the incident signal of element $n$. We assume $\beta_n^{(t)} =1, \forall n \in \mathcal{N}, \forall t \in \mathcal{T}$ without loss of generality \cite{WQQ_TWC19_IRS}. 
By applying receive beamformer ${\bm{m}^{(t)}}$, the obtained signal at BS is 
\begin{align} \label{global_gradient}
	&(\hat{\mathbf{s}}^{(t)})^{\mathrm{T}} = (\bm{m}^{(t)})^{\mathrm{H}} \left(\sum_{k \in \mathcal{K}}\tilde{\bm{h}}_k^{(t)} b_{k}^{(t)} ({\mathbf{s}}^{(t)}_k)^{\mathrm{T}}+\bm{Z}^{(t)}\right)\nonumber \\
	&\!\!\!\!= \!\!\!\!\!\! \underbrace{ \sum_{k \in \mathcal{K}}({\mathbf{s}}^{(t)}_k)^{\mathrm{T}}}_{\text{desired signal:\ }(\overline{\mathbf{s}}^{(t)})^{\mathrm{T}}} \!\!\!\!\! \!\!  + \!\!\underbrace{\sum_{k \in \mathcal{K}}  \Big((\bm{m}^{(t)})^{\mathrm{H}}\tilde{\bm{h}}_k^{(t)} b_{k}^{(t)} \! - \! \! 1 \! \Big) (\mathbf{s}_{k}^{(t)})^{\mathrm{T}} \!\! + (\bm{m}^{(t)})^{\mathrm{H}}\bm{Z}^{(t)}}_{\text{aggregation error:\ }(\bm{\varepsilon}_s^{(t)})^{\mathrm{T}}}.
\end{align}
where $\tilde{\bm{h}}_k^{(t)} =  \bm{h}_{d, k}^{(t)}+\bm{G}^{(t)} \bm{\Theta}^{(t)} \bm{h}_{r, k}^{(t)} $ denotes the superimposed channel from devices to BS, and $\bm{\varepsilon}_s^{(t)}$ is the aggregation error caused by the uplink transmission via AirComp. Hence, the corresponding estimated global gradient $\hat{\bm{g}}^{(t)}$ and the gradient distortion $\bm{\varepsilon}_g^{(t)}$ can be presented as 
\begin{align} \label{global_gradient_1}
	\!\!\hat{\mathbf{g}}^{(t)}\! \! = \!\chi^{(t)}\!(\hat{\mathbf{s}}^{(t)}\!),\bm{\varepsilon}_g^{(t)}\!\! = \! \chi^{(t)}\!(\hat{\mathbf{s}}^{(t)})\! - \! \mathbf{g}^{(t)} \!= \!\chi^{(t)}(\!\mathbf{s}^{(t)}\! \! +\! \bm{\varepsilon}_s^{(t)}\!)\! - \!\chi^{(t)}(\!\mathbf{s}^{(t)}).
\end{align}
Hence, the local model updating at each device becomes
\begin{equation}
\mathbf{w}^{(t+1)}=\mathbf{w}^{(t)}-\alpha^{(t)} \cdot \hat{\mathbf{g}}^{(t)} = \mathbf{w}^{(t)}-\alpha^{(t)} \cdot\left(\overline{\mathbf{g}}^{(t)}+\bm{\varepsilon}_g^{(t)}\right).
\end{equation}

Notice that the aggregated gradient at BS in \eqref{global_gradient_1} is biased due to perturbation caused by the channel fading and noise. In the next section, we analyze the convergence behavior of the AirComp-based FL with perturbed gradients.

\section{Convergence Analysis and Problem Formulation}

In this section, we analyze the convergence behavior of FL. First, we introduce several assumptions on the loss function $F(\bf{w})$ and stochastic gradients to facilitate the convergence analysis, that are commonly made in the existing literature. Then, the generalized optimality gap, which is suitable for generic wireless networks, is derived to characterize the learning efficiency between two arbitrary communication rounds. The obtained optimality gap sheds light on how the imperfect gradient updates affect the convergence of FL. Next, we focus on the IRS-assisted AirComp to provide precise gradients aggregation for FL, and formulate the corresponding performance-oriented design problems.

\subsection{Convergence Analysis}

In this subsection, we present several assumptions which have been widely used in the convergence analysis of FL, see e.g., \cite{SIAM12, SIAM18, JMLR21_Cooperative_SGD}.

\begin{assumption}[Smoothness] \label{As_Smoothness}
The global loss function $F(\mathbf{w})$ is smooth at any point $\mathbf{w} \in \mathbb{R}^{d}$ with constant $L>0$, that is, it is continuously differentiable and the gradient $\nabla F(\mathbf{w})$ is Lipschitz continuous with constant $L$, i.e., 
\begin{equation}
	\|\nabla F(\mathbf{w})-\nabla F(\mathbf{w}^{\prime})\| \leq L\|\mathbf{w}-\mathbf{w}^{\prime}\|, \forall \mathbf{w}, \mathbf{w}^{\prime} \in \mathbb{R}^{d}.
\end{equation}
\end{assumption}

\begin{assumption}[Polyak-{\L}ojasiewicz (PL) condition
\cite{PL_condition_PKDD16}] \label{As_PL}
The global loss function $F(\mathbf{w})$ satisfies the PL condition, i.e., for constant $\mu>0$,
\begin{equation}
\left\|\nabla F(\mathbf{w}^{(t)}) \right\|^2 \geq 2\mu(F(\mathbf{w}^{(t)})-F(\mathbf{w}^*) ),
\end{equation}
where $\mathbf{w}^*$ is the global minimizer of the loss function. 
\end{assumption}

\begin{assumption}[Unbiased estimation] \label{As_Unbias}
The stochastic gradient evaluated on a mini-batch $\hat{\mathcal{D}}_k \subset \mathcal{D}$ and at any point $\mathbf{w}$ is an unbiased estimator of the partial full gradient, i.e. $\mathbb{E}[\mathbf{g}_k^{(t)}]=\nabla F(\mathbf{w}^{(t)})$.
\end{assumption}

%
%

As shown in the following theorem, the convergence behavior for generic wireless networks with perturbed gradients is proposed based on the above assumptions.

\begin{theorem}[Generalized Optimality Gap] \label{Theorem_Gap}
Suppose that an FL procedure satisfies Assumptions 1-3, and the learning rate $\alpha^{(t)} \equiv \alpha$ with $\alpha \leq \frac{1}{\mu}$ and $\alpha \leq \frac{1}{L}$,\footnote{Note that the optimality gap with time-varying learning rate is also given in Appendix A, and the following problem formulation and proposed solution can be readily implemented on the one with varying $\alpha^{(t)}$.} 
for arbitrary $T_2, T_1$ satisfy $ T_2 > T_1$, the optimality gap at the end of $T_2$-th round compared to the one of $T_1$-th is bounded by 
\begin{align} \label{Gap}
&\mathbb{E}\left[F\left(\mathbf{w}^{(T_2+1)}\right)\right]-F(\mathbf{w}^ *) \nonumber \\
&\leq  \left(1- \mu\alpha \right)^{(T_2-T_1)}\left(\mathbb{E}\left[F\left(\mathbf{w}^{(T_1+1)}\right)\right]-F(\mathbf{w}^ *)\right) \nonumber \\
& + \! \! \!\! \sum_{t=T_1+1}^{T_2} \!\! \left(1 \!- \! \mu\alpha \right)^{(T_2-t)}  \bigg\{ \frac{\alpha (1- L \alpha)}{2} \left\|\mathbb{E}\left[\bm{\varepsilon}_g^{(t)}\right]\right\|^{2} \nonumber \\
& +\frac{L \alpha^{2}}{2} \mathbb{E}\left[\left\|\bm{\varepsilon}_g^{(t)}\right\|^{2}\right] + \frac{L \alpha^{2}}{2} \mathbb{E}\left[\left\|\overline{\mathbf{g}}^{(t)}\right\|^{2}\right] \bigg\},
\end{align}
where $\bm{\varepsilon}_g^{(t)} =  \chi^{(t)}(\mathbf{s}^{(t)} + \bm{\varepsilon}_s^{(t)}) - \chi^{(t)}(\mathbf{s}^{(t)})$ is the gradient distortion.
\end{theorem}

{\it{Proof}}. See Appendix A. $\hfill\blacksquare$ 

The presented optimality gap in Theorem $1$ is shown as a weighted sum of distortions $\bm{\varepsilon}_g^{(t)}$ in each communication round. For arbitrary communication round $t_1$, $t_2$ satisfy $T_1 < t_1<t_2 < T_2$, the time-related weights $(1- \mu\alpha)^{(T_2-t_1)} < (1- \mu\alpha)^{(T_2-t_2)}$ due to that the base $(1- \mu\alpha)$ belongs to $(0,1)$ and the exponents $T_2-t_1 > T_2-t_2 \geq 1$. Hence, more communication resources are required to diminish the aggregation error in the later rounds thus minimizing the optimality gap which exhibits the ``\textit{latter-is-better}'' principle. 
Next, we formulate the offline and online problems based on Theorem $1$.

\subsection{Problem Formulation}

The distortion caused by the channel fading and noise have a strong impact on the FL convergence, therefore, we utilize the IRS to configure favorable channels thereby reducing the aggregation error. As we focus on the ultimate performance of FL, the goal is to directly minimize the optimality gap in the final round with the maximum power constraint and long-term energy budget constraint of each device. 
Note that distortion $\bm{\varepsilon}_g^{(t)}$ in \eqref{Gap} is highly related to two aspects, one is the wireless channel, the other is normalization and de-normalization processes. Typically, the transmit symbols in AirComp are normalized into zero-mean unit-variance symbols based on current local gradient statistics \cite{HangLiu_TWC21_FL_IRS, YuanmingShi_TWC21_FL_IRS}. Although \eqref{Gap} can be readily transformed into the case by setting $T_2 = T$, and $T_1 = 0$, one critical issue appears that such a long-term system optimization requires foreseeing the future information, i.e., complete offline channel state information (CSI) and the normalization and de-normalization functions shown in \eqref{aggregating} over the entire FL period (i.e., all the $T$ learning rounds) which is impractical. In this part, we propose two schemes, one is the offline design with lookahead information served as the benchmark, and the other is the online design conducted on each communication round.

\subsubsection{Offline Design With Lookahead Information} We divide the entire FL period into $R \geq 1$ periods, each comprising of $\rho \geq 1$ learning rounds such that $T = R\rho$. The offline framework is based on the $\rho$-rounds lookahead CSI and estimated $\ell_2$-norm bound of local gradients in each round. We assume the CSI in the next $\rho$ rounds is known in advance which is reasonable when the CSI during all the $\rho$-rounds remains static in a slow-fading channel, or the CSI can be precisely predicted via deep learning or other techniques \cite{DeepChannel}. In addition, the $\ell_2$-norm bound of local gradients in the next $\rho$ rounds are estimated by the heuristic methods provided in \cite{JSAC22_Scheduling}.
In that case, the normalization process at each device is conducted as 
\begin{align}
	\mathbf{s}_k^{(t)} = \frac{\mathbf{g}_k^{(t)}}{\sqrt{\gamma^{(t)}} },\ \forall k,
\end{align}
where $\sqrt{\gamma^{(t)}}$ is the estimated $\ell_2$-norm bound according to \cite{AAAI19, JSAC22_Scheduling}.
The de-normalization process at the BS is 
\begin{align}
	\hat{\mathbf{g}}^{(t)} = \frac{\sqrt{\gamma^{(t)}}}{K} \hat{\mathbf{s}}^{(t)} =\overline{\mathbf{g}}^{(t)} + \frac{\sqrt{\gamma^{(t)}}}{K} \bm{\varepsilon}_s^{(t)}.
\end{align}
Hence, the corresponding gradient aggregation error is given by
\begin{align}
	\bm{\varepsilon}_g^{(t)} =  \frac{\sqrt{\gamma^{(t)}}}{K}  \bm{\varepsilon}_s^{(t)},
\end{align}
and we have 
\begin{align} \label{MSE_bound}
	\left\|\mathbb{E}\left[\bm{\varepsilon}_g^{(t)}\right]\right\|^{2} \! \!\!
	&=\!\! \bigg|\frac{1}{K} \! \sum_{k \in \mathcal{K}}  \! \Big((\bm{m}^{(t)})^{\mathrm{H}}\tilde{\bm{h}}_k^{(t)} b_{f,k}^{(t)} \!\! - \! 1 \Big) \bigg|^2 \! \! \|\nabla F(\mathbf{w}^{(t)})\|^{2} \! \nonumber \\
	&\leq \! \bigg|\frac{1}{K} \! \sum_{k \in \mathcal{K}}  \Big((\bm{m}^{(t)})^{\mathrm{H}}\tilde{\bm{h}}_k^{(t)} b_{f,k}^{(t)} \!\! - \!\! 1 \Big) \bigg|^2 \! \! \gamma^{(t)} \! , \! \\ 
	\! \! \mathbb{E}\left[\left\|\bm{\varepsilon}_g^{(t)}\right\|^{2}\right] & \!\! \leq  \!\!\bigg( \!\! \frac{1}{K^2} \! \! \sum_{k \in \mathcal{K}}  \Big|(\bm{m}^{(t)})^{\mathrm{H}}\tilde{\bm{h}}_k^{(t)} b_{f,k}^{(t)} \! - \! 1 \Big|^2 \!\! \! \! + \! \frac{1}{K^2} \|\bm{m}^{(t)}\!\|^{2}d \sigma_z^2 \! \bigg) \gamma^{(t)},
\end{align}
where $b_{f,k}^{(t)}$ denotes the transmit factor of the offline scheme.

According to \eqref{Gap}, the optimality gap at the end of period $r$ compared to that of period $r-1$ is bounded as
\begin{align} \label{gap_r}
\!\!\!\!\!\!&\mathbb{E}\left[\! F\left(\mathbf{w}^{(r\rho+1)}\right)\right]\!\! -\!\! F(\mathbf{w}^ *) 
\!\! \leq \!\! \big(1\!- \! \mu\alpha \big)^{\rho}\!\!\underbrace{\big(\mathbb{E}\!\left[F(\mathbf{w}^{((r-1)*\rho+1)})\big]\!\! - \!F(\mathbf{w}^ *)\right)}_{\text{Optimality gap in the end of period $r-1$}} \nonumber \\
&+ \!\! \!\!\! \! \!\!\underbrace{\sum_{t=(r-1)\rho+1}^{r\rho} \! \! \! \!\! \! \!\!\omega_1^{(t)} \left\|\mathbb{E}\left[\bm{\varepsilon}_g^{(t)}\right]\right\|^{2} \! + \!\! \! \! \!\sum_{t=(r-1)\rho+1}^{r\rho} \! \!  \! \! \!\! \!\!\omega_2^{(t)} \!\!\left(\! \mathbb{E}\left[\left\|\bm{\varepsilon}_g^{(t)}\right\|^{2}\right] \!\!+ \!\mathbb{E}\left[\left\|\overline{\mathbf{g}}^{(t)}\right\|^{2}\right]\right)}_{\Lambda^{(r)}({\{b_k^{(t)}\},\{\bm{\Theta}^{(t)}\},\{\bm{m}^{(t)}}\})}\!, \!\!
\end{align} 
where $\omega_1^{(t)} = \left(1- \mu\alpha \right)^{(T_2-t)} \frac{\alpha (1- L \alpha)}{2}$, and $\omega_2^{(t)} = \left(1- \mu\alpha \right)^{(T_2-t)} \frac{ L \alpha^2}{2}$.
Hence, the ultimate gap at round $T$ (i.e., the end of period $R$) is bounded as
\begin{align} \label{gap_T}
&\mathbb{E}\left[F \!\! \left(\mathbf{w}^{(T+1)}\right)\right] \!   - \! \! F (\mathbf{w}^ *) \leq  \! \left(1\! \! - \! \! \mu\alpha \right)^{T} \! \!\left(\mathbb{E}\left[F\! \left(\mathbf{w}^{(1)}\right)\right]\! \! - \! \!F(\mathbf{w}^ *)\right) \nonumber \\
&\quad \ + \! \sum_{r=1}^R \left(1\!- \! \mu\alpha \right)^{(R-r)\rho} \Lambda^{(r)}({\{b_k^{(t)}\}, \! \{\bm{\Theta}^{(t)}\}, \! \{\bm{m}^{(t)}}\}).
\end{align}

We consider the maximum transmit power on an individual symbol to be ${P}_{k}^{\mathrm{max}}$, thus each edge device is subject to a maximum power constraint $d{P}_{k}^{\mathrm{max} }$ on each communication round, i.e., 
\begin{equation} \label{maximum}
	\! \! \mathbb{E}\left(\left\||b_{f,k}^{(t)}| \mathbf{s}_{k}^{(t)}\right\|^{2}\right) \leq |b_{f,k}^{(t)}|^2 \leq d{P}_{k}^{\mathrm{max} }, \forall k \in \mathcal{K}, \forall t \in \mathcal{T},
\end{equation}
Besides, to reveal the communication resources allocation between different communication rounds, we further consider the average power constraint of each period which equivalently substitutes the long-term energy budget constraint:
\begin{equation} \label{avg}
	\! \! \! \! \!\! \!\! \!\sum_{t=(r-1)\rho+1}^{r\rho} \!\! \!\! \!\! \!\!\mathbb{E} \left( \!\left\||b_{f,k}^{(t)}| \mathbf{s}_{k}^{(t)}\right\|^{2} \! \right) \!\! \leq \!\! \!\! \!\!\sum_{t=(r-1)\rho+1}^{r\rho} \!\! \!\! \!\! |b_{f,k}^{(t)}|^2 \! \leq  \!  \rho d {P}_{k}^{\mathrm{avg} },  \! \forall k \in \mathcal{K}. \!\!
\end{equation} 
By discarding the constant term in \eqref{gap_T}, i.e., the initial optimality gap, the corresponding optimization problem\footnote{The formulated problem can be readily extended to the one including device scheduling, and two kinds of extensions are provided in Appendix B.} is given by
\begin{subequations}
\begin{align}
\! \! \!\! \!\! \!(\mathrm{P1}): \! \! \! \! \! \!\!\! &\mathop{\min}_{{\{b_{f,k}^{(t)}\}, \atop\{\bm{\Theta}^{(t)}\},\{\bm{m}^{(t)}\}}} \!   \! \! \! \!\sum_{r=1}^R \! \left(1\!- \! \mu\alpha \right)^{(R-r)\rho} \! \! \Lambda^{(r)}({\{b_{f,k}^{(t)}\},\{\bm{\Theta}^{(t)}\},\{\bm{m}^{(t)}}\}) \! \!\! \\
&\quad {\rm s.t.} \ \ |b_{f,k}^{(t)}|^2  \leq d{P}_{k}^{\mathrm{max}}, \forall k \in \mathcal{K}, \forall t \in \mathcal{T}, \label{power_constrain} \\
&\quad \ \ \ \ \sum_{t=(r-1)\rho+1}^{r\rho}  \!\!\!\!\! |b_{f,k}^{(t)}|^2  \! \leq \! \rho  d {P}_{k}^{\mathrm{avg}}\!\!, \forall k \in \mathcal{K}, \!\forall r \in [1,R], \label{power_avg} \!\\
&\quad \ \ \ \ \ 0\leq\theta_n^{(t)}\leq2\pi, \forall n \in \{1,\dots,N\}, \forall t \in \mathcal{T} .\label{continuous_constraint}
\end{align}
\end{subequations}
where the objective function is a weighted sum of the optimality gap in each communication round, and the latter rounds have larger weights. Furthermore, $(\mathrm{P1})$ can be divided into $R$ subproblems since both the objective function and the constraints are independent among different periods. Each subproblem designs the resource allocation in one period in which the communication resources can be flexibly allocated. 
In practice, the CSI and the estimated gradient bounds may have errors, thereby the energy consumption needs to be corrected at the end of each period $r$. Note that the presented offline design scheme is served as a benchmark for the online scheme shown below.

\subsubsection{Online Design via Lyapunov Technique} 
$(\mathrm{P}1)$ needs the $\rho$-rounds lookahead CSI as well as the estimated gradient norm bound to allocate the communication resources. However, the CSI varies rapidly within a period when facing a fast-varying channel, and the channel prediction techniques may induce additional errors. 
In addition, the estimated $\ell_2$-norm bound of local gradients may induce additional perturbation to the obtained optimality gap. Hence, we utilize the Lyapunov optimization framework \cite{Lyapunov_Book} to construct a virtual energy deficit queue ${e}_k(t)$ for each device $k$ to guide the power allocation over sequential communication rounds without foreseeing the future. 
According to \cite{HangLiu_TWC21_FL_IRS, YuanmingShi_TWC21_FL_IRS}, the normalization process at each device is conducted as 
\begin{align}
	\mathbf{s}_k^{(t)} = \frac{\mathbf{g}_k^{(t)}- \frac{1}{K} \sum_{k \in \mathcal{K}} \xi_k^{(t)}}{\frac{1}{K}\sqrt{\sum_{k \in \mathcal{K}} (\iota_k^{(t)})^2}},\ \forall k,
\end{align}
where $\xi_k^{(t)}$, $(\iota_k^{(t)})^2$ is the local gradient statistics (i.e., the means and variances) in current round $t$ that are computed as
\begin{align}
	\xi_k^{(t)} = \frac{1}{D} \sum_{d=1}^D g_{k,d}^{(t)},\ (\iota_k^{(t)})^2 = \frac{1}{D} \sum_{d=1}^D (g_{k,d}^{(t)} - \xi_k^{(t)})^2,\ \forall k.
\end{align}
The de-normalization process at the BS is
\begin{align}
	&\hat{\mathbf{g}}^{(t)} = \frac{1}{K} \bigg( \frac{\sqrt{\sum_{k \in \mathcal{K}} (\iota_k^{(t)})^2}}{K}\hat{\mathbf{s}}^{(t)} + \sum_{k \in \mathcal{K}} \xi_k^{(t)} \bigg),
\end{align}
thereby, the corresponding gradient aggregation error is
\begin{align}
	\bm{\varepsilon}_g^{(t)} = \hat{\mathbf{g}}^{(t)} - \overline{\mathbf{g}}^{(t)} = \frac{\sqrt{\sum_{k \in \mathcal{K}} (\iota_k^{(t)})^2}}{K^2}  \bm{\varepsilon}_s^{(t)},
\end{align}
%
and we have
\begin{align}
	&\left\|\mathbb{E}\left[\bm{\varepsilon}_g^{(t)}\right]\right\|^{2} = 0,\nonumber \\
	&\mathbb{E}\left[\left\|\bm{\varepsilon}_g^{(t)}\right\|^{2}\right] \!\! \leq \!\! \frac{d\sum_{k \in \mathcal{K}} (\iota_k^{(t)})^2}{K^4} \! \bigg( \! \sum_{k \in \mathcal{K}}  \Big|(\bm{m}^{(t)})^{\mathrm{H}} \tilde{\bm{h}}_k^{(t)} b_{k}^{(t)} \!\! - \! 1 \Big|^2  \!\! \! \! + \!\! \|\bm{m}^{(t)}\|^{2} \sigma_z^2 \bigg).
\end{align}

Consider the maximum transmit power on an individual symbol to be ${P}_{k}^{\mathrm{max}}$ as the offline scheme, thereby the maximum power constraint of the online scheme is given by
\begin{align}
	|b_{k}^{(t)}|^2  \leq  {P}_{k}^{\mathrm{max}}, \forall k \in \mathcal{K}, \forall t \in \mathcal{T}.
\end{align}
The virtual energy queue of device $k$ starts with $e_k(t) \geq 0, \forall k$, and is updated at the end of each round $t$ as
\begin{equation} \label{queue}
	e_k(t+1) = \max\{e_k(t) +  d|b_k^{(t)}|^2 - d{P}_{k}^{\mathrm{avg}}, 0 \},
\end{equation}
where $d|b_{k}^{(t)}|^2$ is the power consumption on communication round $t$. Hence, $e_k(t)$ indicates the deviation of the current energy consumption of device $k$ from its long-term energy constraint $\rho d {P}_{k}^{\mathrm{avg}}$. Let $\bm{e}(t) =  \{ e_1(t), e_2(t), \ldots, e_K(t) \}$ collect the energy deficit queues for all devices. 
The constructed optimization problem of communication round $t$ is shown as a weighted sum of the per-round optimality gap and the energy consumption at the devices, i.e.,
\begin{subequations}
\begin{align}
\!\!\!\! (\mathrm{P2}): \!\!\!\!\!\!\! &\mathop{\min}
\limits_{\scriptstyle
{\{b_k^{(t)}\}, \{\bm{\Theta}^{(t)}\},
\atop \scriptstyle \{\bm{m}^{(t)}\}}
} \! \! \! \! \! \! \!\! V_r\left(1\!- \! \mu\alpha \right)^{(R-r)\rho} \!\omega_2^{(t)} \mathbb{E}\left[\left\|\bm{\varepsilon}_g^{(t)}\right\|^{2}\right] \! \!+ \! \sum_{k \in \mathcal{K}} d e_k(t) |b_{k}^{(t)}|^2 \label{P2_obj} \!\! \! \\
&\qquad \  \mathrm{s.t.} \ \ |b_{k}^{(t)}|^2  \leq  {P}_{k}^{\mathrm{max}}, \forall k \in \mathcal{K}, \forall t \in \mathcal{T}, \\
&\qquad \qquad \  0 \! \leq\theta_n^{(t)}\leq2\pi, \forall n \in \{1,\dots,N\},\! \forall t \! \in \! \!\mathcal{T},
\end{align}
\end{subequations}
where parameter $V_r \geq 0$ is represented as an importance weight to adjust the emphasis on the objective function (i.e., optimality gap minimization) in different communication rounds. Notice that if $e_k(t)$ increases in round $t$, then minimizing energy consumption is more critical in round $t+1$. Intuitively, when the energy queue is stable, the constraint in \eqref{power_avg} is satisfied. Although problems $(\mathrm{P2})$ and $(\mathrm{P1})$ are not equivalent due to the different objective functions, constraints, and even different normalization/de-normalization processes, the following theorem unveils that the results of $(\mathrm{P2})$ are comparable to $(\mathrm{P1})$ and are within a bounded deviation from the optimal results of the omniscient $(\mathrm{P1})$ with precise future information.

\begin{theorem}Assume that the omniscient $(\mathrm{P1})$ is an offline scheme that obtains precise CSI and local gradient statistics at the beginning of each period and conducts the normalization/de-normalization as the one in the online scheme. $\mathrm{(P2)}$ can achieve a performance-backlog tradeoff of $[\mathcal{O}(1/V), \mathcal{O}(\sqrt{V})]$ with compared to omniscient $(\mathrm{P1})$. We use $G^*$ to represent the obtained optimality gap via the optimal offline solution and $G^\dag$ to represent obtained the optimality gap via  $\mathrm{(P2)}$. Specifically, the FL performance is $\mathcal{O}(1/V)$-optimal, which is bounded by
\begin{align} \label{Lyapunov_theorem1}
	 \sum_{t=1}^{T} G_t^\dag \leq  \sum_{t=1}^{T} G_t^* + \sum_{r=1}^{R-1} \frac{C_r}{V_r},
\end{align}
where $ C_r = \rho C_e + \sum_{t=r\rho+1}^{(r+1)\rho}\sum_{k \in \mathcal{K}} E_{\max}\big((t-1)E_{\max} + e_k(r\rho+1)\big) $, and  $ E_{\max} = \max_{k,t} (d|b_k^{(t)}|^2 - d{P}_{k}^{\mathrm{avg}})$. Besides, the energy consumption of each device is $\mathcal{O}(\sqrt{V})$-bounded as 
\begin{align} \label{Lyapunov_theorem2}
	 &\sum_{t=1}^T \left( d|b_k^{(t)}|^2 - d{P}_{k}^{\mathrm{avg}} \right) \nonumber \\ 
	 &\leq \sum_{r=1}^{R-1} \Bigg( \sqrt{2\bigg(C_r + V_r \sum_{t=r\rho+1}^{(r+1)\rho} G_t^*\bigg)} - e_k(r\rho+1) \Bigg).
\end{align}
\end{theorem}

{\it{Proof}}. See Appendix C. $\hfill\blacksquare$

\section{Proposed Algorithms for Offline and Online Design}

In the previous section, we present an offline optimization problem based on the available lookahead information and an online optimization problem via the Lyapunov technique without foreseeing the future. It is observed that $(\mathrm{P1})$ can be divided into $R$ subproblems, each of which corresponds to one period, and $\mathrm{(P2)}$ is an online design problem within each communication round. Note that although variables are highly coupled in the objective function, all the constraints corresponding to each variable are uncoupled with others in both problems. This thus motivates us to apply the BCD method to solve them efficiently by properly partitioning the optimization variables into different blocks. One can observe that, with any given transmit factor, the optimization procedures for receiver $\bm{m}^{(t)}$ and IRS phase shift $\bm{\Theta}^{(t)}$ of $\mathrm{(P2)}$ is a simplification of $\mathrm{(P1)}$ with drop the term related to $\left\|\mathbb{E}\left[\bm{\varepsilon}_g^{(t)}\right]\right\|^{2}$. Besides, there exist similarities in the optimization procedures for power allocation of $(\mathrm{P1})$ and $(\mathrm{P2})$. Therefore, we propose the optimization algorithms for the two problems from the perspective of different variable blocks.

\vspace{-10pt}
\subsection{Power Allocation}

For any given $\{\bm{\Theta}^{(t)}\}$ and $\{\bm{m}^{(t)}\}$,  the transmit factors need to be precisely allocated among different communication rounds to minimize the ultimate optimality gap. Recalling $\mathrm{(P1)}$ and $\mathrm{(P2)}$, we can obtain the following remark.

\begin{remark} 
	 To minimize the objective functions of $\mathrm{(P1)}$ and $\mathrm{(P2)}$, each device $k$ in communication round $t$ always adjusts the phase of its transmit factor for phase compensation with respect to its equivalent fading channel $ \bar{{h}}_k^{(t)} = (\bm{m}^{(t)})^{\mathrm{H}}\tilde{\bm{h}}_k^{(t)}$, i.e., $  \mathrm{arg}(b_{f,k}^{(t)}) = \mathrm{arg}(b_k^{(t)}) = \mathrm{arg}\big((\bar{{h}}_k^{(t)})^{-1}\big) $. 
\end{remark}

In that case, only the magnitudes of $\{ b_{f,k}^{(t)} \}$ have effects on the value of the objective function. 
By fixing $\mathrm{arg}(b_{f,k}^{(t)}) = \mathrm{arg}\big((\bar{{h}}_k^{(t)})^{-1}\big) $, and letting $\bar{b}_{f,k}^{(t)} = |b_{f,k}^{(t)}|, \forall k \in \mathcal{K}$, the power allocation problem of $\mathrm{(P1)}$ over one dedicated period can be expressed in the real form as
\vspace{-5pt} 
\begin{subequations}
\begin{align} \label{P_transmit}
&(\mathrm{P1.1}):\mathop{\min}_{\{\bar{b}_{f,k}^{(t)}\}}\ \ \!\!\! \sum_{t=(r-1)\rho+1}^{r\rho} \! \! \!\! \! \! \frac{\gamma^{(t)} \omega_1^{(t)}}{K^2}    \left( \sum_{k\in\mathcal{K}} |\bar{{h}}_k^{(t)}| \bar{b}_{f,k}^{(t)} \! - \!  K\right)^2 \! \! \! \! \nonumber \\
&\qquad \qquad \qquad \!\!\!\! + \sum_{t=(r-1)\rho}^{r\rho}  \!\!\!\! \frac{\gamma^{(t)} \omega_2^{(t)} }{K^2} \sum_{k \in \mathcal{K}} \left( |\bar{{h}}_k^{(t)}| \bar{b}_{f,k}^{(t)} - 1 \right)^2 \\
&\quad {\rm s.t.} 
(\bar{b}_{f,k}^{(t)})^2  \leq \! d {P}_{k}^{\max}, \! \forall t \! \in \! [(r\!-\!1)\rho+1,r\rho], \forall k \in \mathcal{K}, \\
&\qquad \sum_{t=(r-1)\rho+1}^{r\rho} \! \! \! \! \! \! (\bar{b}_{f,k}^{(t)})^2 \! \leq \rho d {P}_{k}^{\mathrm{avg}}, \forall k \in \mathcal{K}. \label{sum_power}
\end{align}
\end{subequations}
Note that $(\mathrm{P1.1})$ is a convex quadratic optimization problem which can be optimally solved by standard convex optimization techniques such as the interior point method or alternating direction method of multipliers (ADMM). Instead, we next resort to the Lagrange duality method to derive the structured optimal solution for problem $(\mathrm{P1.1})$ to gain engineering insights. Let $\lambda_k^*$ denote the optimal dual variable associated with the $k$-th constraint in \eqref{sum_power}. By adopting the first-order optimality condition, the optimal solution of $\bar{b}_k^{(t)}$ is given by
\vspace{-5pt}
\begin{align}\label{power_allocation}
	\!\!(\bar{b}_{f,k}^{(t)})^* \! \! = \! \!\operatorname{min} \Bigg\{ \!\frac{ \! \gamma^{(t)} \omega_1^{(t)} \! (K \! \! - \! \sum_{i\neq k } \! |\bar{{h}}_i^{(t)}| \bar{b}_{i}^{(t)}) \! + \!  \gamma^{(t)} \omega_2^{(t)}}
	 {\big(\gamma^{(t)} \omega_1^{(t)} +  \gamma^{(t)} \omega_2^{(t)} \big)|\bar{{h}}_k^{(t)}| + \frac{K^2}{|\bar{{h}}_k^{(t)}|} \lambda_k^*}, \sqrt{ d {P}_{k}^{\max}} \! \Bigg\},
\end{align}
where the optimal dual variables $\{\lambda_k^*\}$ can be obtained through the subgradient-based methods as shown in \cite{XiaowenCao_JSAC22}. 

Consider the special case that the average power budget is not less than the maximum power at all devices, i.e., ${P}_{k}^{\mathrm{avg}} \geq {P}_{k}^{\max}, \forall k \in \mathcal{K}$. Then, the average power constraints are not activated, and the dual variables are equal to zero (i.e., $\lambda_k^*=0, \forall k$). The proposed optimal power allocation strategy is reduced to the isolated optimization across different communication rounds, and exhibits a threshold-based structure: all the devices $k\in \mathcal{K}' $ that cannot compensate their channel fading, i.e., satisfying $|\bar{{h}}_i^{(t)}|  \sqrt{ d {P}_{k}^{\max}} - 1 < 0$, should transmit signals with the largest power; otherwise, the devices $k\notin \mathcal{K}' $ that satisfying $|\bar{{h}}_i^{(t)}|  \sqrt{ d {P}_{k}^{\max}} - 1 \geq 0$, should enlarge their transmit power moderately (overcompensate their channel) to minimize the optimality gap exacerbated by the deep faded channel of the devices $k \in \mathcal{K}'$. As for the general case that the average power budgets are less than the maximum power at all devices, i.e., ${P}_{k}^{\mathrm{avg}} < {P}_{k}^{\max}, \forall k \in \mathcal{K}$, the dual variables in \eqref{power_allocation} are larger than zero (i.e., $\lambda_k^*>0$). From \eqref{power_allocation}, one can observe that the power allocation is closely related to the chronological order and the current channel condition. Larger power will be allocated to the devices in the later communication rounds and/or with better channel conditions.

The power control of $\mathrm{(P2)}$ in one dedicated  communication round is a convex optimization problem presented as
\begin{subequations}
\begin{align}
&\mathop{\min}_{{\{b_k^{(t)}\}}} V_r\left(1\!- \! \mu\alpha \right)^{(R-r)\rho} \!\omega_2^{(t)} \frac{\sum (\iota_k^{(t)})^2}{K^4} \sum_{k \in \mathcal{K}}  \Big||\bar{{h}}_k^{(t)}| \bar{b}_k^{(t)} \! - \! 1 \Big|^2 \! \! \nonumber \\ 
& \qquad \quad+ \! \sum_{k \in \mathcal{K}}  e_k(t) |b_{k}^{(t)}|^2 \label{P2_obj} \!\! \! \\
&\qquad \  \mathrm{s.t.} \ \ |b_{k}^{(t)}|^2  \leq  {P}_{k}^{\mathrm{max}}, \forall k \in \mathcal{K}, \forall t \in \mathcal{T}, 
\end{align}
\end{subequations}
and the optimal solution is given by
\vspace{-5pt}
\begin{align} \label{power_allocation2}
	\!(\bar{b}_{k}^{(t)})^{\dag} \!\! =  \! \operatorname{min} \Bigg\{  \! \frac{ 1 } { \!|\bar{{h}}_k^{(t)}\!|\! + \!\frac{e_k(t)K^4}{V_r\left(1\!- \! \mu\alpha \right)^{(R-r)\rho} \!\omega_2^{(t)}\! \sum (\iota_k^{(t)})^2  |\bar{{h}}_k^{(t)}|}}, \sqrt{ {P}_{k}^{\max}} \Bigg\}.
\end{align}
It can be observed that the power allocation is closely related to the current energy queue $e_k(t)$ and the parameter $V_r$ which acted as an importance weight to balance the emphasis on the optimality gap and energy consumption minimization, in addition to the chronological order and the current channel condition. We can obtain a similar result as $\mathrm{(P1)}$ that larger power will be allocated to the devices in the later communication rounds and/or with better channel conditions. Besides, the allocated power will be diminished when facing a higher energy queue $e_k(t)$ and/or less importance weight $V_r$ on the optimality gap in order to satisfy the long-term energy constraint.

\vspace{-5pt}
\subsection{Receiver Design}

For any given power allocation and phase shift of IRS, the optimization problems with respect to the receive beamformers $\{\bm{m}^{(t)}\}$ of $(\mathrm{P1})$ can be separated among different rounds. The corresponding subproblem of $(\mathrm{P1})$ with omitting the time index can be expressed as
\vspace{-5pt}
\begin{align}\label{problem_receiver}
(\mathrm{P1.2}):\mathop{\min}_{\bm{m}}&\ \omega_1  \Big| \bm{m}^{\mathrm{H}} \sum_{k\in\mathcal{K}}\tilde{\bm{h}}_{k} b_{f,k}  -  K\Big|^2 \nonumber \\
&+  \omega_2 \!\Big( \sum_{k \in \mathcal{K}} \! \Big|  \bm{m}^{\mathrm{H}}\tilde{\bm{h}}_k b_{f,k} \! \!- \!1 \Big|^2 \! \!  + d \sigma^2_z \|\bm{m}\|^2 \!\Big). \!
\end{align}
Since problem $(\mathrm{P1.2})$ is an unconstrained convex problem, by exploiting the first-order optimality condition, the closed-form solution for $\bm{m}$ is given by
\vspace{-5pt}
\begin{align} \label{opt_receiver}
	\bm{m}^* \! = &\bigg( \!  \omega_1  \Big(\! \sum_{k\in\mathcal{K}}\tilde{\bm{h}}_k b_{f,k} \Big)^2 \! + \! \omega_2  \sum_{k\in\mathcal{K}} |b_{f,k}|^2 \tilde{\bm{h}}_k\tilde{\bm{h}}_k^{\mathrm{H}} + \omega_2 d \sigma_z^2 \bm{I} \bigg)^{-1} \nonumber \\
	 & \Big( \big(K  \omega_1  + \omega_2  \big) \sum_{k\in\mathcal{K}} \tilde{\bm{h}}_k b_{f,k} \Big).
\end{align}
The optimal receiver \eqref{opt_receiver} of the proposed offline design scheme is shown as a weighted summed structure, with the corresponding weights being the ones of the bias term and the MSE term in \eqref{Gap}. Similarly, it can be obtained that the optimal receiver of the online design scheme is given by $\left(\sum_{k \in \mathcal{K}}\left|b_{k}\right|^{2} \tilde{\boldsymbol{h}}_{k} \tilde{\boldsymbol{h}}_{k}^{\mathrm{H}}+\sigma_z^{2} \mathbf{I}\right)^{-1} \left(\sum_{k\in\mathcal{K}} \tilde{\bm{h}}_k b_k\right)$ that targets at computing the summation of local gradients.

\vspace{-10pt}
\subsection{Phase Shift Optimization of the IRS}

For any given power allocation and the receiver beamformer, the optimization problem with respect to the phase shift $\{\bm{\Theta}^{(t)}\}$ of $(\mathrm{P1})$ can be separated among different communication rounds. Note that optimization procedures for $\{\bm{\Theta}^{(t)}\}$ of $(\mathrm{P2})$ is part of the subproblem of $(\mathrm{P1})$ that is only related to the MSE term $\mathbb{E}\Big[\big\|\bm{\varepsilon}_g^{(t)}\big\|^{2}\Big]$. Hence, we analyze the corresponding subproblem of $(\mathrm{P1})$ and obtain the solutions of the two problems. By multiplying by a constant term $K^2$, the subproblem with respect to the phase shift $\{\bm{\Theta}\}$ of $(\mathrm{P1})$ is given by
\begin{subequations}
\begin{align}
(\mathrm{P1.3}): \! \mathop{\min}_{\bm{\Theta}}& \; \omega_1  \Big| \bm{m}^{\mathrm{H}} \sum_{k\in\mathcal{K}}\tilde{\bm{h}}_k b_{f,k} \!  -  \! K\Big|^2 \! \! + \! \omega_2 \sum_{k \in \mathcal{K}} \Big|  \bm{m}^{\mathrm{H}}\tilde{\bm{h}}_k b_{f,k} \! - \! 1 \Big|^2 \\
	{\rm s.t.}&\ \  0\leq\theta_n\leq2\pi, \forall n \in \{1,\dots,N\}. \label{P4_theta}
\end{align}
\end{subequations}
Since $(\mathrm{P1.3})$ is a non-convex optimization problem due to the per-phase constraints in \eqref{P4_theta}. We first apply the semidefinite relaxation (SDR) strategy to relax $(\mathrm{P1.3})$ into a semidefinite programming (SDP) problem. To reduce the computational complexity, we further propose an element-vise optimization approach to achieve near-optimal performance. Besides, it can be readily extended to the practical scenario that with discrete phase shifts constraint.

\subsubsection{SDR Approach}
Let $\bm{v} = [e^{\jmath\theta_1}, \ldots, e^{\jmath\theta_N} ]^{\mathrm{H}}$, $(\mathrm{P1.3})$ can be transformed to a convex optimation problem with norm-one constraint. Furthermore, the well-known SDR technique can be applied to convert it to a convex SDP problem \cite{WQQ_TWC19_IRS, ZhaoWZZ21, ZhiquanLuo_SDR} which can be solved efficiently in polynomial time by existing convex optimization solvers such as CVX. Besides, additional steps such as Gaussian randomization \cite{ZhiquanLuo_SDR} need to be applied to extract a suboptimal solution when the optimal solution to the SDP problem is obtained with a higher rank.

\subsubsection{Low-complexity Suboptimal Solution}

Although a satisfactory solution to $(\mathrm{P1.3})$ can be obtained through the SDR approach, the lifted optimization variable induces huge computation complexity. Solving a series of high-dimensional SDP problems drastically increases the computational burden. Besides, the additional steps to extract a suboptimal solution are often with intolerable computation consumption, especially when facing an IRS with large-scale elements that are common in practice. Furthermore, the obtained continuous phase shifts are practically difficult to implement due to hardware limitations. Generally, the discrete phase shifts optimization problem is shown as an integer linear program problem that can be optimally solved via the branch-and-bound with the worst-case exponential complexity \cite{WQQ_Discrete}. To reduce the computational complexity and provide the near-optimal solution for the IRS with discrete phase shifts constraint, we successively refine the phase shift of each element until converges.  It is noted that the phase shifts of all elements are fully separable in the constraint, and only coupled in the objective function. Hence, we can successively optimize the phase shift of each element until converges. 
For a given $n \in \mathcal{N}$, by fixing the others, the objective function in $(\mathrm{P1.3})$ is linear with respect to $e^{\jmath\theta_n}$, which is written as
\begin{align} \label{f_theta_n}
&\omega_1 \bigg(\bm{v}^{\mathrm{H}} \underbrace{\bm{\psi}\bm{\psi}^{\mathrm{H}}}_{\bm{\Psi}'} \bm{v} +  2\Re\{\bm{v}^{\mathrm{H}}\underbrace{\bm{\psi}\zeta^{\mathrm{H}}}_{\zeta'}\}  + |\zeta|^2\bigg)  \nonumber \\
& \quad + \omega_2  \sum_{k \in \mathcal{K}} \bigg(\bm{v}^{\mathrm{H}}\underbrace{\bm{\phi}_k\bm{\phi}_k^{\mathrm{H}}}_{\bm{\Phi}'_k} \bm{v} + 2\Re\{\bm{v}^{\mathrm{H}}\underbrace{\bm{\phi}_k\varphi_k^{\mathrm{H}}}_{\varphi'_k}\} + |\varphi_k|^2  \bigg) \nonumber \\
&=\!\!  \omega_1 \bigg(\sum_{l\neq n}^N \sum_{i\neq n}^N \bm{\Psi}'(l,i)e^{\jmath(\theta_l-\theta_i)} + 2\Re\{e^{\jmath\theta_n}q^{(1)}_n\} + C' \bigg) \nonumber \\
&+\!\!  \omega_2 \bigg( \! \! \sum_{k=1}^{K}\!  \sum_{l \neq n}^{N} \! \sum_{i \neq n}^{N} \!\bm{\Phi}'_k(l, i) e^{\jmath\left(\theta_{l}\!-\!\theta_{i}\right)} \!\! +  \!\! 2 \Re\Big\{e^{\jmath \theta_{n}} \sum_{k=1}^{K} q^{(2)}_{k, n}\Big\}\! + \! \sum_{k=1}^{K} \! C_{k} \! \bigg),
\end{align}
where $\bm{\psi} = \operatorname{diag}(\bm{m}^{\mathrm{H}} \bm{G})\bm{H}_r \bm{b}$, $\zeta = \bm{m}^{\mathrm{H}} \bm{H}_{d}\bm{b} - K$, $\bm{\phi}_k = \operatorname{diag}(\bm{m}^H\bm{G})\bm{h}_{r,k} b_{f,k}$, $\varphi_k = \bm{m}^H\bm{h}_{d, k} b_{f,k} -~1$, and $\bm{H}_r, \bm{H}_{d}, \bm{b}$ collect the channel and the power allocation of devices, respectively. Besides, the constant terms in \eqref{f_theta_n} are given by
\begin{align}
	&q^{(1)}_n = \sum_{l\neq n}^N \bm{\Psi}'(n,l)e^{-\jmath\theta_l} +  \zeta'(n) = |q^{(1)}_n| e^{\jmath \nu_{n}}, \\
	&C' =  \bm{\Psi}'(n,n) + 2\Re\{\sum_{l\neq n}^N e^{\jmath \theta_l} \zeta'(l) \} + | \zeta'|^2, \\
	&\!\! \sum_{k=1}^{K} \! q^{(2)}_{k, n} \! = \! \sum_{k=1}^{K} \! \!\left(\sum_{l \neq n}^{N} \bm{\Phi}'(n, l) e^{-\jmath \theta_{l}}+\varphi'_{k}(n)\right) \! \! = \! \! |q^{(2)}_n| e^{\jmath \varsigma_{n}}, \\
	&C_{k}=\bm{\Phi}'(n, n)+2 \Re\left\{\sum_{l \neq n}^{N} e^{\jmath \theta_{\ell}} \varphi'_{k}(l) \right\}+ |\varphi'_{k}|^{2}.
\end{align}
By leveraging the trigonometric identities, the part related to element $n$ can be equivalently transformed to
\begin{align}
	& \omega_1  \Re\{e^{\jmath \theta_n}q^{(1)}_n\} +  \omega_2 \Re\left\{e^{\jmath \theta_{n}} \sum_{k=1}^{K} q^{(2)}_{k, n}\right\} \nonumber \\ 
	=& \omega_1 |q^{(1)}_n| \sin(\theta_n+\nu_{n}+\frac{\pi}{2})  +  \omega_2 |q^{(2)}_n|\sin(\theta_n+\varsigma_{n}+\frac{\pi}{2})\nonumber \\	
	=& \hat{c}_n \sin(\theta_n + \hat{\theta}_n),
\end{align}
where $\hat{\theta}_n = \mathrm{atan} \frac{\omega_2 |q^{(2)}_n|\sin(\varsigma_{n}-\nu_{n})}{\omega_1|q^{(1)}_n| + \omega_2 |q^{(2)}_n|\cos(\varsigma_{n}-\nu_{n})} \in [-\frac{\pi}{2}, \frac{\pi}{2}]$ if $\omega_1|q^{(1)}_n| + \omega_2 |q^{(2)}_n|\cos(\varsigma_{n}-\nu_{n})>0$, otherwise $\hat{\theta}_n = \pi + \mathrm{atan} \frac{\omega_2 |q^{(2)}_n|\sin(\varsigma_{n}-\nu_{n})}{\omega_1|q^{(1)}_n| + \omega_2 |q^{(2)}_n|\cos(\varsigma_{n}-\nu_{n})}  \in [\frac{\pi}{2}, \frac{3\pi}{2}]$. 

Hence, the optimal phase shift of $(\mathrm{P1.3})$ for the element $n$ is 
\begin{equation}
	\theta_{n}^{*} = \frac{3}{2} \pi -\hat{\theta}_n.
\end{equation}
Besides, the optimal phase shift of the element $n$ to minimize the objective function of $(\mathrm{P2})$ is $\theta_{n}^{*} = \arg \min_{\theta_n} \sin(\theta_n+\varsigma_{n}+\frac{\pi}{2})$. For ease of practical implementation, we consider that the phase shift at each element of the IRS can take only a finite number of discrete values. Let $\delta$ denote the number of bits used to indicate the number of phase shift levels. For simplicity, we assume that such discrete phase shift values are obtained by uniformly quantizing the interval $[0,2\pi)$. Thus, the set of discrete phase shift values at each element is given by $
	\mathcal{S}=\{0, \Delta \theta, \ldots,(2^{\delta}-1) \Delta \theta\}$ with $\Delta \theta = 2\pi/2^{\delta}$. Hence, the optimal configuration with finite quantization bit is $\bar{\theta}_{n}^{*}=\arg \min _{\theta \in \mathcal{S}}\left|\theta-\theta_{n}^{*}\right|$.
With successively setting the phase shifts of all elements based on the above, the objective value will be non-increasing over the iterations until converges.

\textbf{Overall Algorithm and Computational Complexity Analysis}: Based on the provided solutions to the above three subproblems, an efficient BCD algorithm is proposed, where the IRS beamforming vector and transceiver are alternately optimized until convergence is achieved. Note that the objective value of problem $(\mathrm{P1})$ and $(\mathrm{P2})$ is bounded and non-decreasing by alternately optimizing $\{\bm{m}^{(t)}\}$, $\{b_k^{(t)}\}$ and $\{\bm{\Theta}^{(t)}\}$. Besides, the proposed BCD algorithm is guaranteed to converge to the stationary points of the two problems since the variables are only coupled in the objective functions.
The mainly computational complexity of the proposed BCD algorithm lies in solving the sub-problems $(\mathrm{P1.1})$, $(\mathrm{P1.2})$, and $(\mathrm{P1.3})$. Specifically, the corresponding computational complexity is given by $\mathcal{O}((\rho K )^{3.5})$, $\mathcal{O}(\rho M)$ and $\mathcal{O}(\rho N I'_{iter})$, where $I'_{iter}$ denotes the number of iterations required in the element-vise optimization of discrete phase shifts. Therefore, the total complexity of the BCD algorithm is $\mathcal{O}\Big( \big( (\rho K )^{3.5}+ \rho (N I'_{iter} + M) \big) I_{iter} \Big)$, where $I_{iter}$ denotes the number of iterations required to reach convergence of the objective function.

$(\mathrm{P2})$ is an online system design problem that is operated in each communication round. The corresponding complexity of the BCD algorithm for $(\mathrm{P2})$ is $\mathcal{O}\Big( \big( K+ N I'_{iter}+ M\big) I_{iter} \Big)$.
From Theorem 2, we can obtain $(\mathrm{P2})$ is only a bounded substitution of omniscient $(\mathrm{P1})$. $V_r$ is utilized to control the trade-off between the sizes of the queue backlogs and the objective function value.
It is noted that an increasing $V_r$ will strengthen the emphasis on the optimality gap minimization rather than power minimization in the later rounds, therefore the ultimate performance can be promoted. Besides, if the $e_k(r\rho+1)$ is initialized to $0$ as in the existing literature \cite{OptimalityGap_TWC21}, $(\mathrm{P2})$ is reduced to an optimality gap minimization problem in the initial rounds without virtual energy constraint. In that case, the obtained power allocation in the initial is most like to exceed the average power constraint, which obviously deviates from the optimal allocation. Therefore, the value of $V_r$ and $e_k(r\rho+1)$ are highly related to the FL performance, and we will show this phenomenon in the next section.

\section{Simulation Results}

In this section, we conduct extensive simulations to validate the performance of the proposed performance-oriented long-term design approach for IRS-assisted FL. Simulations are operated on the MNIST dataset \cite{MNIST} and the CIFAR-10 dataset \cite{Krizhevsky09}. First, the convergence behavior of the proposed optimality gap minimization algorithm was presented. Then, we analyze the test accuracy under different parameters, e.g., $\rho$ which determines the available future information which is highly related to the power allocation among different communication rounds, and the number of IRS elements $N$ which determines the capability to configure the wireless channel. Besides, we compare the performance of offline and online design approaches under different settings, and provide the power allocation over the entire FL procedure to depict the \textit{later-is-better} principle.

\subsection{Simulation Setup}

We consider a three-dimensional coordinate system, where the BS and the IRS are respectively located at $(0, 0, 30)$ and $(0, 50, 20)$ meters. In addition, the edge devices are randomly distributed in the circle region centered at $(50, 40, 0)$ with the radius equals to $20$ meters. The wireless channels from the devices to the BS over different communication rounds follow i.i.d. Rayleigh fading, and the channels from IRS to BS and devices follow i.i.d. Rician fading. The path loss model under consideration is $L(d) = T_0 (d/d_0)^\alpha $, where $T_0 = -30 $ dB is the path loss at reference distance $d_0= 1$ meter, $d$ is the signal distance, and $\alpha$ is the path loss exponent. The path loss exponents for the BS-device link, the BS-IRS link, and the IRS-device link are set to $3.5$, $2.2$, and $2.5$, respectively. Other parameters are set as follows: ${P}_{k}^{\mathrm{max} } = 20$ dBm and ${P}_{k}^{\mathrm{avg} } = 17$ dBm, $\sigma_z^2=-80$ dBm.

\subsection{Performance Evaluation}

\begin{figure}[h]
\centering
\includegraphics[width=2.8in]{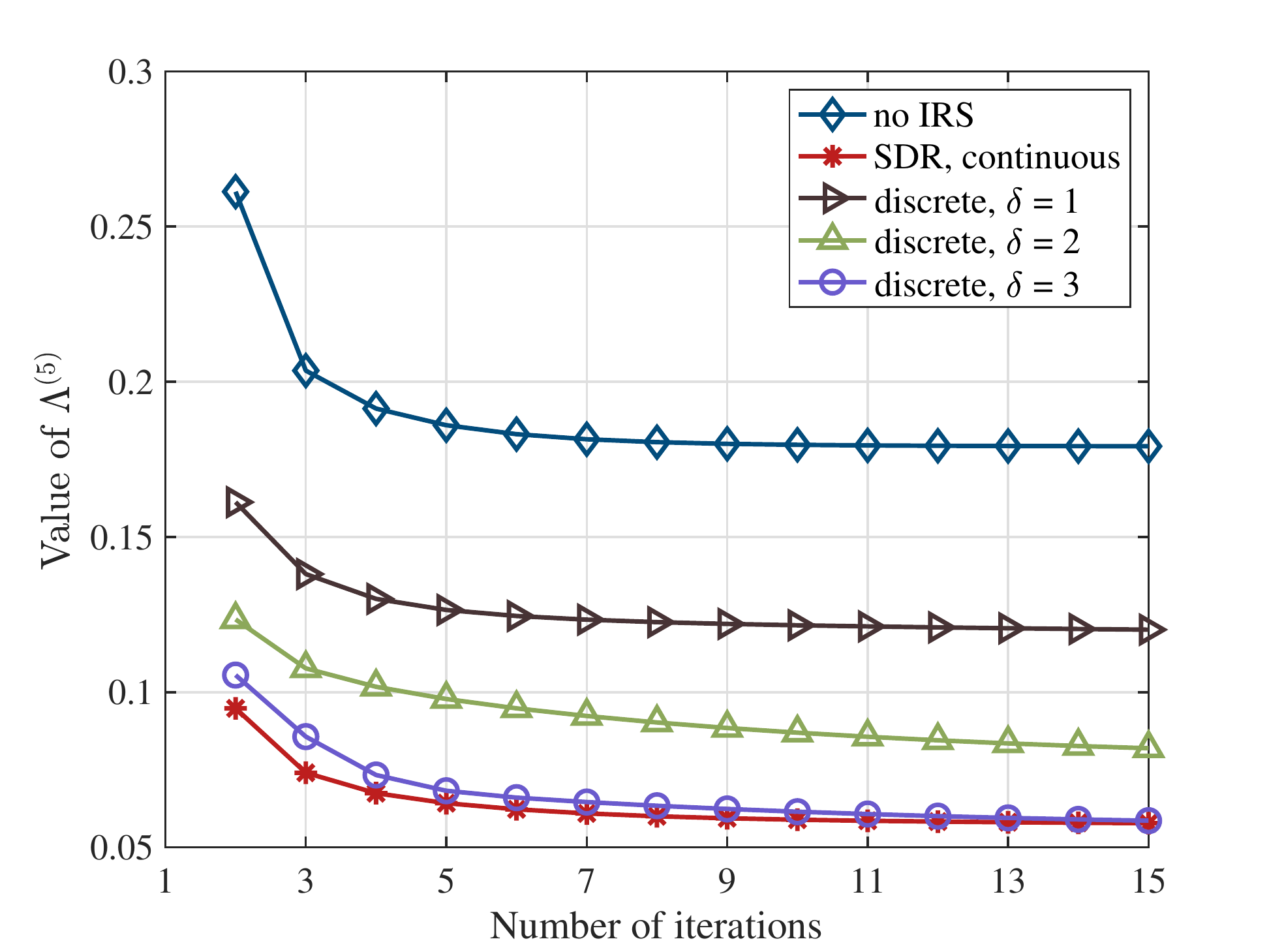}
\caption{Convergence behavior of the proposed algorithm.}
\label{fig_convergence}
\end{figure}
\vspace{-5pt}


First, the convergence behavior of the proposed offline scheme on the MNIST dataset \cite{MNIST} is shown in Fig. \ref{fig_convergence}. We set $T=100$, $R=10$, and $\rho=10$, the convergence behavior of the proposed optimality gap minimization algorithm conducted on the period $5$ (i.e., $40$ to $50$ communication rounds) is presented as an example. Besides, parameters in the assumptions are set as $\mu = 0.2, L=10$, respectively  \cite{Mohammad_TWC22}. The number of devices is fixed to $K = 20$. For the case without IRS, the phase shifts matrix is fixed to $\mathbf{\Theta} = \mathbf{0} $. For the case with IRS, the number of the IRS elements is fixed to $N = 40$, and the number of antennas equipped at BS is set to $M= 5$. It is shown that the IRS can significantly diminish the optimality gap even with a finite configuration range such as the quantization bit equals to $1$. Besides, the performance of IRS with a quantization level of $3$ can achieve nearly the same performance as the one with continuous phase shifts.

Then, the IRS-assisted FL was deployed for handwritten recognition on the MNIST dataset \cite{MNIST} and image classification on the CIFAR-10 dataset \cite{Krizhevsky09}. We implement a $3$-layer CNN as the recognition model for the MNIST dataset, which consists of an input layer, a final softmax output layer, and a midterm convolution layer with max pooling. The local batch size at each edge device is set to $B = 64$, and the learning rate $\alpha_t$ is fixed to $0.005$. Besides, three convolution layers followed by max-pooling layers are developed for classification on the CIFAR-10 dataset, and other parameters are set as $\mu=1, L=5, B = 256, \alpha_t = 0.01$ \cite{BlindFL}. We compare the learning performance with the following two benchmark schemes:
\begin{itemize}
	\item \textbf{Optimal aggregation}: The wireless channels are ideal, thus the server can aggregate precise local gradients without the aggregation error presented in \eqref{global_gradient}.
	\item \textbf{Isolated Design}: Variables are isolated optimized in each communication round as the existing works to minimize the aggregation error, hence the maximum transmit power constraint becomes $P_{\rm{avg}}$.
\end{itemize} 

\begin{figure*} [t!] 
	\centering
	\subfloat[\label{fig_M_offline}Accuracy of the offline design.]{\hspace{-5mm}
		\includegraphics[scale=0.3]{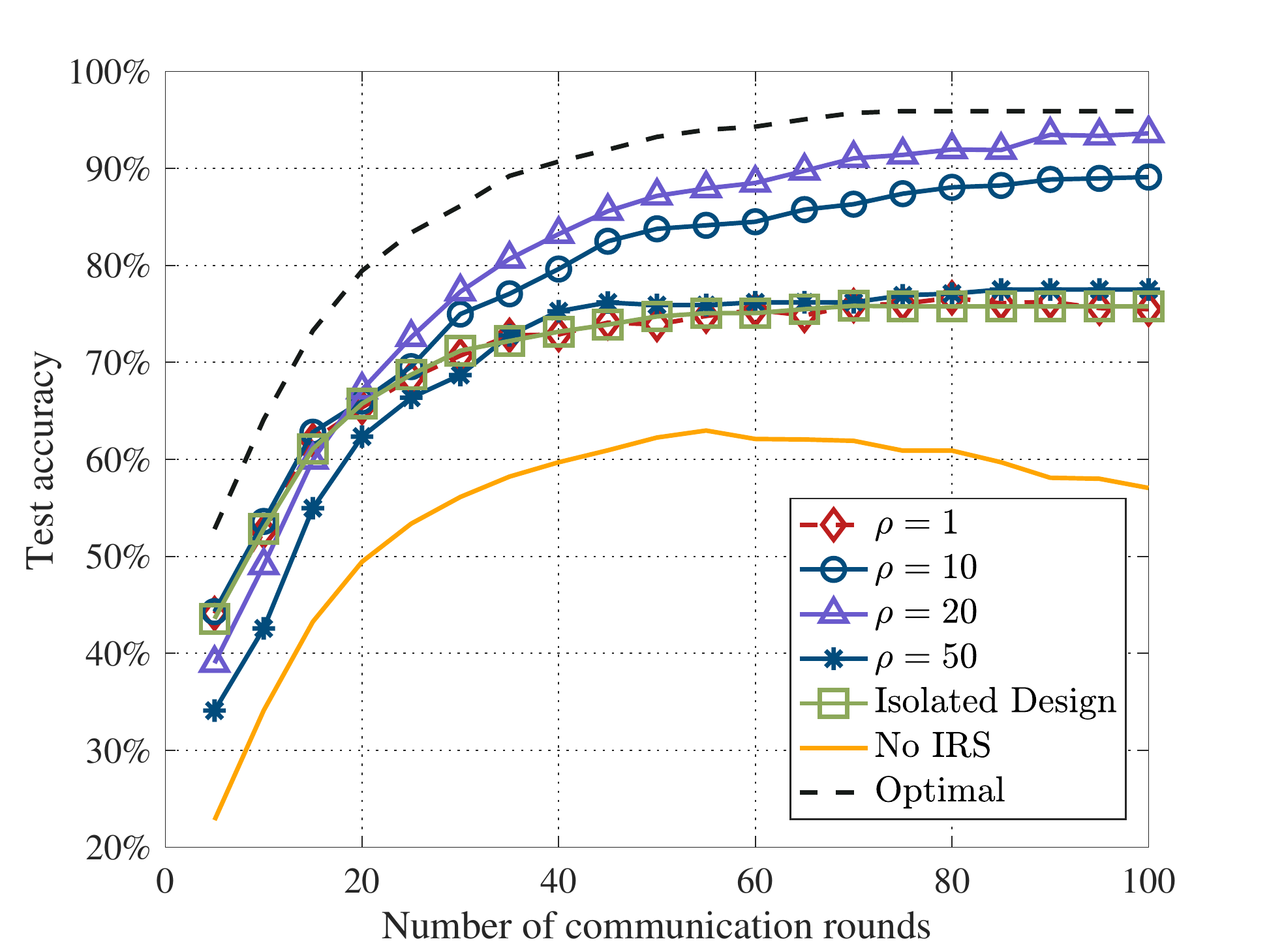}}
	\subfloat[\label{fig_M_online}Accuracy of the online design.]{\hspace{-5mm}
		\includegraphics[scale=0.3]{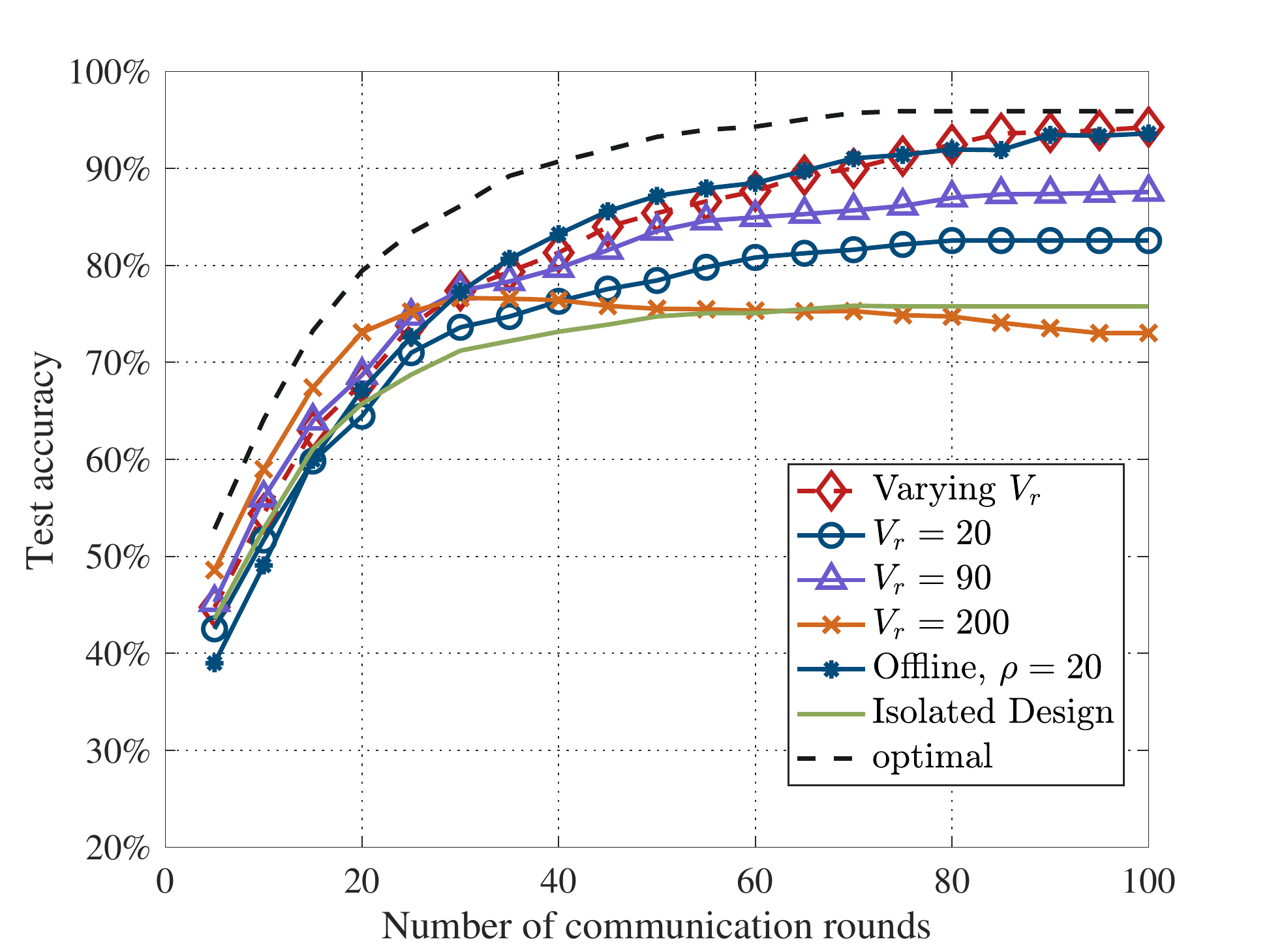}} 
	\subfloat[\label{fig_M_power}Power allocation of the online design.]{\hspace{-5.5mm}
		\includegraphics[scale=0.3]{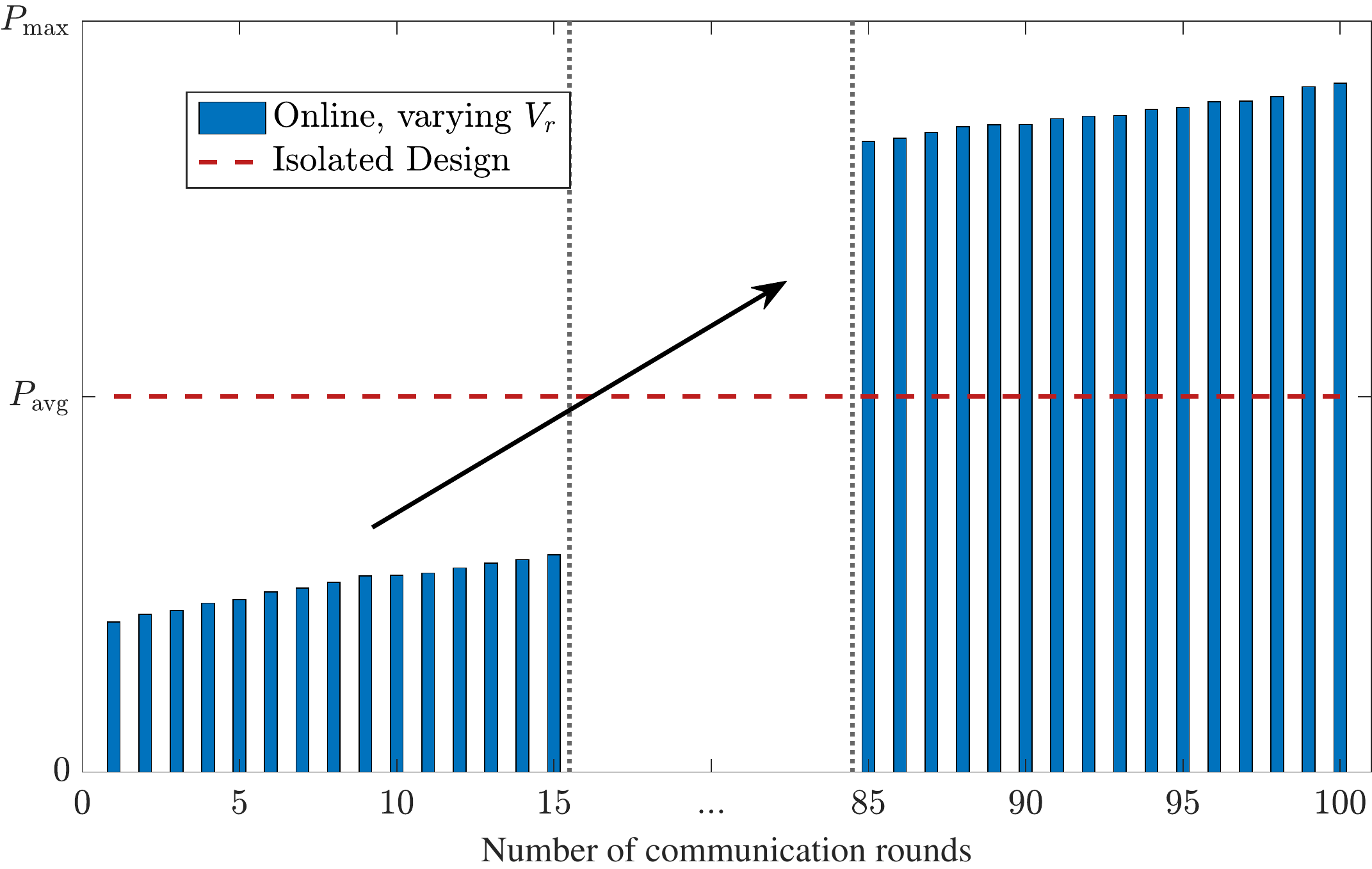}} 
	\caption{Performance of the offline/online design scheme on MINST dataset.}
\vspace{-10pt} 
\label{fig_M_accuracy}
\end{figure*}

\begin{figure*} [t!]
\vspace{-10pt} 
	\centering
	\subfloat[\label{fig_C_offline}Test accuracy of the offline design scheme.]{\hspace{-0mm}
		\includegraphics[scale=0.35]{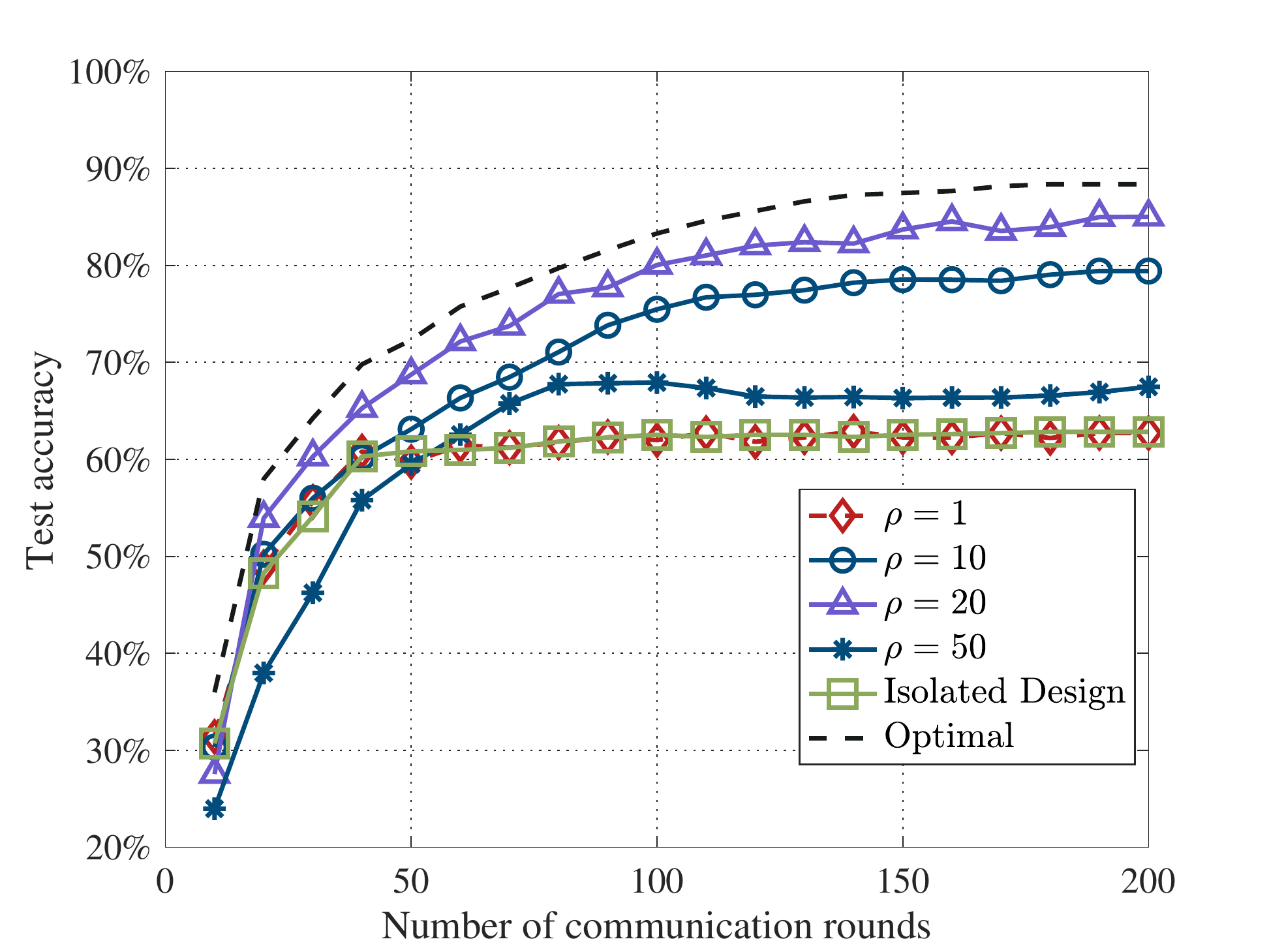}} 
	\subfloat[\label{fig_C_online}Test accuracy of the online design scheme.]{\hspace{-0mm}
		\includegraphics[scale=0.35]{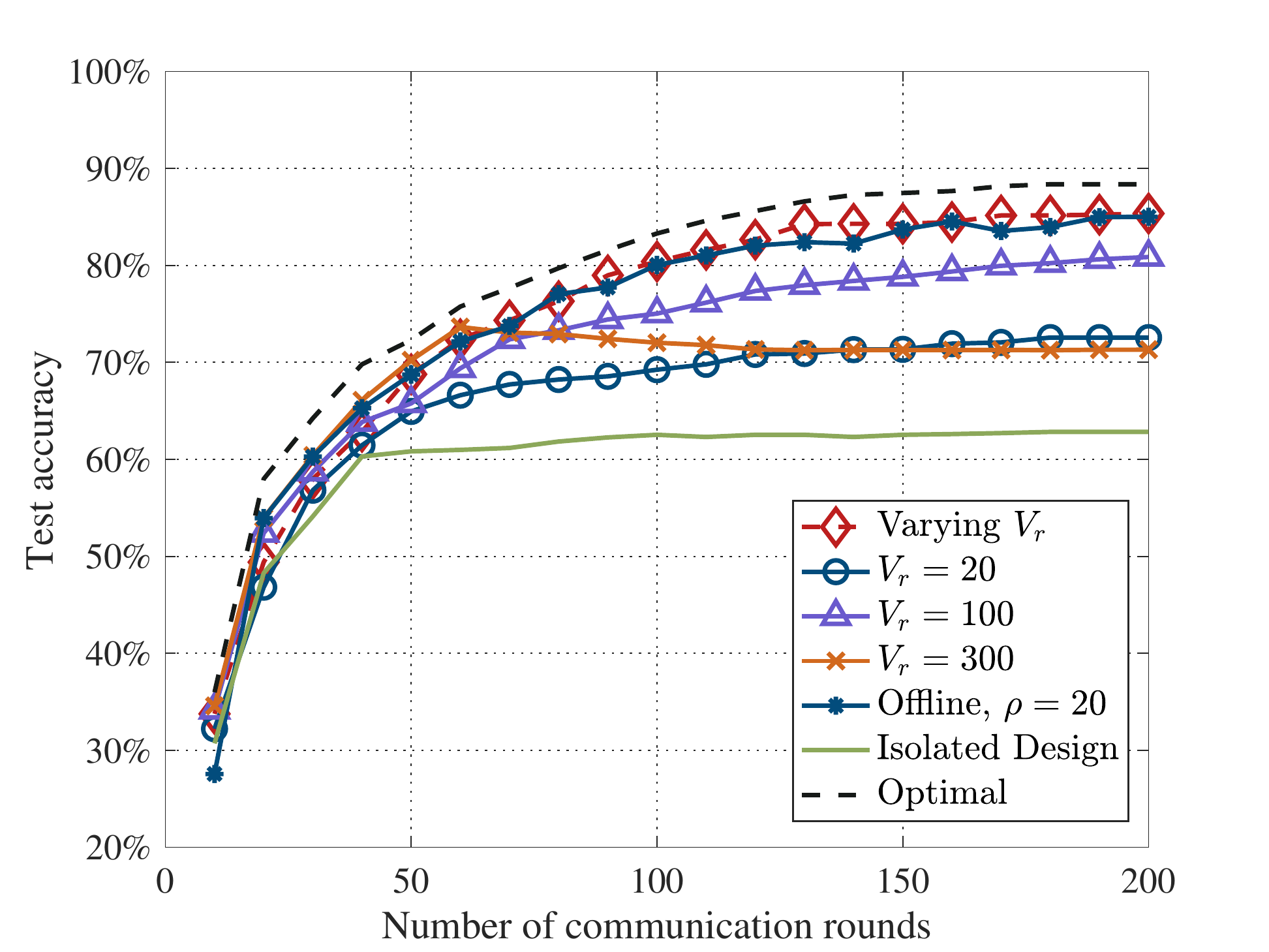}} 
	\caption{Performance of the offline/online design scheme on CIFAR-10 dataset.}
\vspace{-10pt} 
\label{fig_C_accuracy}
\end{figure*}

Fig. \ref{fig_M_offline} and \ref{fig_C_offline} present the test accuracy of the offline scheme on the two datasets with different $\rho$. 
It is observed that the offline scheme with $\rho=1$ has the same performance as the conventional isolated design approach since the offline scheme degenerates to the isolated design with no future information available.
Besides, the performance of $\rho=10$ and $\rho=20$ are inferior to the isolated design approach initially, but it speeds up and eventually outperforms in the later rounds. This is due to the fact that the contributions of aggregation errors to the optimality gap are distinct at different communication rounds, which cannot be captured by the conventional isolated design scheme. 
Note that the system optimization is operated based on each $\rho$ rounds lookahead information, therefore, the available information is closely related to the system performance. It is observed that the offline scheme with $\rho=20$ outperforms the one with $\rho=10$. Moreover, one can observe that the performance declines in the initial as the $\rho$ goes, but rises and exceeds in the later rounds. These observations are consistent with the convergence analysis in the former sections which further demonstrate the necessity and efficiency of such a performance-oriented design approach. The ultimate performance of the proposed algorithm monotonically increases with a moderate $\rho$ since more communication rounds are integrated optimized. 
However, it is observed that the offline scheme with $\rho=50$ has the worst performance compared to $\rho=10$ and $\rho=20$ since the lookahead CSI and the estimated gradient bounds in the distant rounds may have errors thus degrading the system performance. 
Besides, the performance of the FL without the assistance of the IRS deteriorates heavily as the number of communication rounds increases. This is due to the model aggregation error in the latter rounds deteriorating the learning performance greater than the beginning.

Fig. \ref{fig_M_online} and \ref{fig_C_online} depict the test accuracy of the online learning scheme on the two datasets. The FL procedures contain $T=100$ rounds and $T=200$ rounds on the MNIST and the CIFAR-10 dataset, respectively. We choose varying $V_r$ sequences with $V_r = 20\sqrt{(10r+1)}$ for both the MNIST and CIFAR-10 dataset. Besides, the benchmarks with fixed $V_r$ are set as $V_r=20$, $V_r=90$, $V_r=200$ and $V_r=20$, $V_r=100$, $V_r=300$ for the two datasets, respectively. The energy queue $e_k(r\rho+1), \forall r$ are initialized as $0.2d{P}_{k}^{\mathrm{avg}}$. It can be observed that the online design scheme can achieve satisfying performance compared to the offline without foreseeing the future, which can be regarded as a practical and promising approach for FL system design. Besides, Fig. \ref{fig_M_online} and \ref{fig_C_online} validate that the online design approach with varying $V_r$ can obtain higher test accuracy than the fixed $V_r$. The online design schemes with low $V_r$ are more conservative to meet the long-term energy constraints while at the expense of sacrificing the system performance. On the contrary, the online design scheme with high $V_r$ obtains near-optimal performance in the beginning, but it results in extreme energy consumption which is destructive to the latter rounds. It can be observed from the figures that an increasing $V_r$ queue leads to larger power allocation in the latter rounds in FL, thus better performance can be achieved. Furthermore, we present the simulation results of the online scheme on the CIFAR-10 dataset with different parameters in Table \ref{table_2}. It is observed that larger $\rho$ leads to better performance under different $V$. 
Compared it with Fig. \ref{fig_C_online}, we can observed that the online scheme with appropriated $V$ can enhance its performance by enlarge $\rho$ since more communication rounds can be incorporated in the Lyapunov framework, while large $\rho$ deteriorates the performance of the offline scheme as shown in Fig. \ref{fig_C_offline}.

\begin{table}[]
	\centering
	\caption{Performance of the online scheme with different $\rho$}
	\vspace{-5pt}
	\label{table_2}
	\begin{tabular}{c | c  c  c c c}
		\toprule[1pt]
		\midrule
		
		& $\rho=10$   & $\rho=20$ & $\rho=40$ & $\rho=100$  & $\rho=200$ \\ \midrule 
		
		$V=20$   & $70.72\%$ & $72.56\%$     &$74.14\%$  &$75.29\%$   &$77.30\%$   \\ \midrule
		
		$V=100$   & $78.24\%$      & $80.86\%$      &$82.23\%$   &$83.17\%$  &$84.27\%$ \\  \midrule
		
		$V=300$   & $70.91\%$     & $73.30\%$   &$75.43\%$   &$77.90\%$   & $79.29\%$ \\  
		\bottomrule[1pt]
	\end{tabular}
	\vspace{-5pt}
\end{table}

\begin{table}[] 
	\caption{Performance of offline/online scheme under different power sequence}
	\vspace{-5pt}
	\label{table_1}
	\centering
	\setlength{\tabcolsep}{1.5mm}{
	\begin{tabular}{c | c  c  c}
		\toprule[1pt]
		\midrule
		& Descending power&  Equal  power & Proposed algorithm \\ \midrule 
		Offline scheme   & $64.05\%$       &$76.87\%$   & $95.10\%$   \\ \midrule
		Online scheme   & $62.48\%$     &$76.87\%$      &$95.46\%$     \\ 
		\bottomrule[1pt]
	\end{tabular} }
\vspace{-10pt}
\end{table}

\begin{figure*} [t] 
	\centering
	\subfloat[\label{fig_vsN}Test accuracy vs. $N$.]{\hspace{-0mm}
		\includegraphics[scale=0.35]{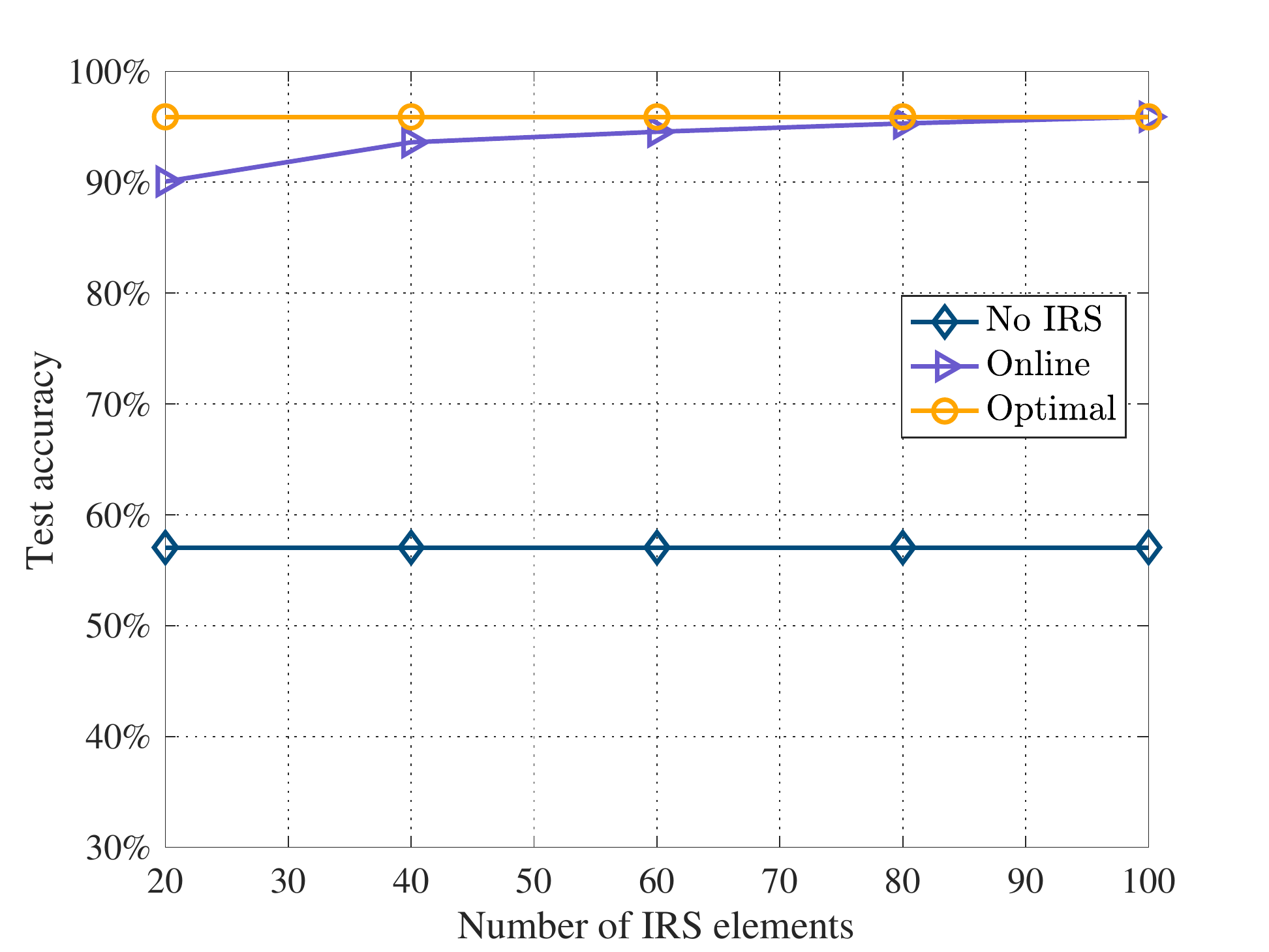}} 
	\subfloat[\label{fig_vsK}Test accuracy vs. $K$.]{\hspace{-0mm}
		\includegraphics[scale=0.35]{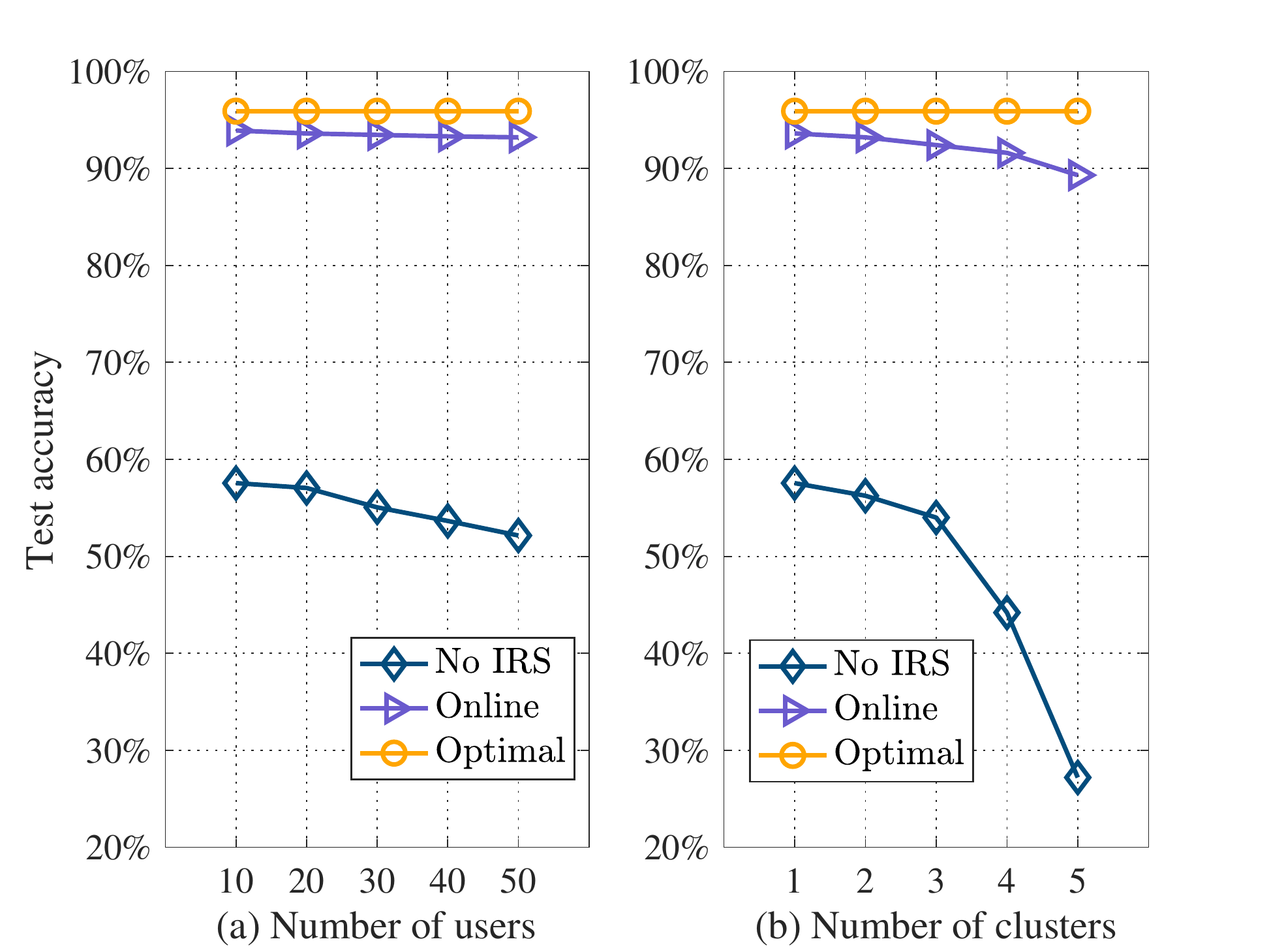}} 
	\caption{Test accuracy on MNIST dataset versus channel parameters.}
	\vspace{-10pt} 
	\label{fig_vsNK}
\end{figure*}

In order to further depict the \textit{later-is-better} theory, we present the power allocation for a specific device of the online scheme on the MNIST dataset in Fig. \ref{fig_M_power}. The simulation is conducted under a static channel, therefore, it can be readily obtained and observed that the transmit power of a fixed device in conventional isolated design schemes remains the same over all the communication rounds. However, the long-term design schemes have a near-monotonic in the transmit power over time.   Furthermore, we test the performance of descending and equal power queue on the MNIST dataset with $\rho=20$ on a stationary channel. We first inverse the obtained power queues of the proposed offline and online schemes to construct the descending power sequences, and then optimize other variables. The equal power queue can be obtained from the conventional isolated design approach since the channel remains static. The simulation results are shown in Table \ref{table_1}. It is observed that both the offline and online schemes with descending power sequence have extremely poor performance compared with the proposed algorithm and the equal power queue. Such an observation further validates the ``\textit{latter-is-better}'' principle. By combing Fig. \ref{fig_M_offline}, \ref{fig_M_online}, \ref{fig_M_power} and Table \ref{table_1}, we can obtain that the system performance is enhanced by allocating large transmit power to the later communication rounds.

Finally, we investigate the test accuracy versus the number of IRS elements\footnote{With assuming that the signal reflected by IRS two or more times is negligible and thus ignored, our problem can be readily extended to the multiple IRSs case. In addition, our proposed algorithm is also applicable to the multiple IRSs case without any modifications.} and the number of users on the MNIST dataset. As shown in Fig. \ref{fig_vsN}, the test accuracy of the proposed algorithm is monotonically increased with $N$ until converges to the optimal. By comparing Fig. \ref{fig_vsN} to Fig. \ref{fig_M_online}, we can observe that a large-scale IRS can diminish the performance deterioration due to the lack of lookahead information. Hence, when the CSI is varying or can not be precisely predicted, e.g., undergoes a fast-varying channel, the IRS with more elements can configure a more favorable communication link thus obtaining higher test accuracy in FL. The performance between FL with IRS and the one without IRS is further compared with the varying numbers of devices. Devices are distributed in one cluster in the setting of Fig. \ref{fig_vsK}(a) and are distributed in various clusters in the setting of Fig. \ref{fig_vsK}(b). Specifically, in the setting of multiple clusters, devices are randomly distributed in $5$ circle regions with each radius equal to $20$ meters, and the centers are located from $(50, 40, 0)$ to $(130, 40, 0)$ with the coordinates of the first dimension incremented by $20$ in sequence. From Fig. \ref{fig_vsK}(a), it is observed that the test accuracy decays slightly as the number of devices increases even without the help of IRS, while the one with IRS can still obtain satisfactory test accuracy with tiny performance loss. This result verifies that AirComp is an efficient data aggregation approach for model aggregation in FL which is tolerant of the number of accessed devices. Then redirecting to Fig. \ref{fig_vsK}(b), one can observe that the performance of the one with or without IRS decays as the number of clusters increases. This is due to that the performance of AirComp is determined by the devices with poor channels. Fortunately, the FL system can still maintain satisfactory performance with the help of IRSs which can configure the deep faded channel to a favorable one.
\vspace{-10pt}

\section{Conclusion}

In this paper, we have proposed a performance-oriented long-term design framework based on the optimality gap for IRS-assisted FL systems. By analyzing the convergence behavior of FL, both offline and online design schemes were established. We adopted the BCD method to tackle the highly-intractable problems. Simulation results have demonstrated that such a long-term design scheme can precisely allocate the resources to different communication rounds, hence achieving a higher test accuracy and faster convergence behavior in FL than the conventional isolated design approach. Besides, we have confirmed that the online design approach based on the Lyapunov framework can achieve satisfactory performance without foreseeing the future which can be regarded as a practical and promising approach for FL system. Further, it has been seen that the employed IRS can assist AirComp in providing precise model aggregation in FL, especially when the lookahead information is limited or devices are widely distributed. The relationship between communication accuracy and learning performance presented in the paper has been rigorously determined via mathematical proof and simulation verification, which have revealed a \textit{later-is-better} principle that can provide significant guidance for the FL system design.


\vspace{-10pt}
\section*{Appendix A: \textsc{Proof of Theorem 1}}
According to Assumption \ref{As_Smoothness}, we can obtain that
\begin{align}
&F\left(\mathbf{w}^{(t+1)}\right)-F\left(\mathbf{w}^{(t)}\right) \nonumber \\
\leq& \! \left\langle \! \nabla F\! \left(\mathbf{w}^{(t)}\right)\!,\! \left(\mathbf{w}^{(t+1)} \! - \! \mathbf{w}^{(t)} \!\right) \! \right\rangle \! + \!\frac{L}{2} \left\|\mathbf{w}^{(t+1)} \!- \mathbf{w}^{(t)}\right\|^{2} \nonumber \\
=&\! - \! \alpha^{(t)} \!\!\left\langle \!  \nabla F \! \left( \! \mathbf{w}^{(t)\!}\right) \! \!, \! \! \left(\overline{\mathbf{g}}^{(t)} \!\! + \!\boldsymbol{\varepsilon}^{(t)}_g\right) \! \! \right\rangle \!\! + \!\! \frac{L (\alpha^{(t)} )^{2}}{2} \!\!\left\|\overline{\mathbf{g}}^{(t)} \! + \!\boldsymbol{\varepsilon}^{(t)}_g\right\|^{2}\!\!\!\!. \!
\label{Appedix_A1}
\end{align}


With $\alpha^{(t)} \leq \frac{1}{L}$, by taking expectation with respect to the stochastic gradient and the aggregation error on both sides of \eqref{Appedix_A1}, we have
\begin{align}
&\!\!\!\!\!\!\!\!\!\!\mathbb{E}\left[F\left(\mathbf{w}^{(t+1)}\right)-F\left(\mathbf{w}^{(t)}\right)\right]
\leq  \frac{L (\alpha^{(t)} )^{2}}{2} \mathbb{E}\left[\left\|\overline{\mathbf{g}}^{(t)} \! \! + \!\boldsymbol{\varepsilon}^{(t)}_g\right\|^{2}\right] \! \!  \nonumber \\
&\! \!-\!\alpha^{(t)}  \mathbb{E} \left[ \left\langle \nabla F\left(\mathbf{w}^{(t)}\right)\! ,\!  \left(\overline{\mathbf{g}}^{(t)} \! \! + \!\boldsymbol{\varepsilon}^{(t)}_g\right) \right\rangle \right] \nonumber  \\
\overset{a}{=}&-\alpha^{(t)}  \left\|\nabla F\left(\mathbf{w}^{(t)}\right)\right\|^{2}-\alpha_t \left\langle \nabla F\left(\mathbf{w}^{(t)}\right), \mathbb{E}\left[\boldsymbol{\varepsilon}^{(t)}_g\right]\right\rangle \nonumber \\
&+\frac{L (\alpha^{(t)} )^{2}}{2} \mathbb{E}\left[\left\|\overline{\mathbf{g}}^{(t)}\right\|^{2}\right] \!\!+ \!\! L (\alpha^{(t)} )^{2} \left\langle \nabla F\left(\mathbf{w}^{(t)}\right), \mathbb{E}\left[\boldsymbol{\varepsilon}^{(t)}_g\right]\right\rangle \nonumber \\
&+\frac{L (\alpha^{(t)} )^{2}}{2} \mathbb{E}\left[\left\|\boldsymbol{\varepsilon}^{(t)}_g\right\|^{2}\right] \nonumber \\
\!\!\overset{b}{\leq}& \!\! - \!\!\alpha^{(t)}  \left\|\nabla F\big(\!\mathbf{w}^{(t)}\big)\right\|^{2}  \!\!\!+\!\!    \frac{L (\alpha^{(t)} )^{2}}{2} \!\! \left( \!\mathbb{E}\left[\left\|\overline{\mathbf{g}}^{(t)}\right\|^{2}\right]  \!\!+\!   \mathbb{E}\left[\left\|\boldsymbol{\varepsilon}^{(t)}_g\right\|^{2}\right] \right) \!  \nonumber \\
& + \!\frac{\alpha^{(t)} (1 - L  \alpha^{(t)} )}{2} \!\! \left(\!\left\|\nabla F\left(\mathbf{w}^{(t)}\right)\right\|^{2}  \!\!+ \Big \|\mathbb{E}\left[\boldsymbol{\varepsilon}^{(t)}_g\right] \Big \|^{2} \right),
\end{align} 
where $a$ is due to Assumption \ref{As_Unbias}, $b$ is due to the Cauchy-Schwarz inequality and the inequality of arithmetic and geometric means: $ \pm \bm{x}_{1}^{T} \bm{x}_{2} \leq \left\|\bm{x}_{1}\right\| \left\|\bm{x}_{2}\right\| \leq \frac{\left\|\bm{x}_{1}\right\|^{2}}{2}+\frac{\left\|\bm{x}_{2}\right\|^{2}}{2}$.

According to Assumption \ref{As_PL}, and with $\alpha^{(t)} \leq \frac{1}{\mu}, \alpha^{(t)} \leq \frac{1}{L}$, we have 
\begin{align} \label{appendix_A_t}
	\!\!&\!\!\!\!\mathbb{E}\left[F\left(\mathbf{w}^{(t+1)}\right)\right]-F(\mathbf{w}^{*}) \nonumber \\
	  &\!\!\!\!\leq \! (1 \! - \! \mu\alpha^{(t)})\! \left(\! \mathbb{E}\left[F(\mathbf{w}^{(t)})\right] \! -\! F(\mathbf{w}^ *)\right) + \frac{L (\alpha^{(t)} )^{2}}{2} \mathbb{E}\left[\left\|\overline{\mathbf{g}}^{(t)}\right\|^{2}\right] \nonumber \\
	  &+ \! \frac{\alpha^{(t)} (1- L \alpha^{(t)})}{2} \Big\|\mathbb{E}\big[\bm{\varepsilon}^{(t)}_g\big]\!\Big\|^{2} +\frac{L (\alpha^{(t)})^{2}}{2} \mathbb{E}\left[\left\|\boldsymbol{\varepsilon}^{(t)}_g\right\|^{2}\right] \! \! .
\end{align} 
Recursively applying \eqref{appendix_A_t}, finally we obtain that for $\forall T_2 > T_1$,
\begin{align} 
\label{appendix_gap}
&\mathbb{E}\left[F\left(\mathbf{w}^{(T_2+1)}\right)\right]-F(\mathbf{w}^ *) \nonumber \\
\leq & \prod_{t=T_1+1}^{T_2} \left(1- \mu\alpha^{(t)} \right)\left(\mathbb{E}\left[F\left(\mathbf{w}^{(T_1+1)}\right)\right]-F(\mathbf{w}^ *)\right) \nonumber \\
+& \! \! \! \! \sum_{t=T_1+1}^{T_2} \frac{\prod_{t=T_1+1}^{T_2}\left(1- \mu\alpha^{(t)} \right)}{\left(1- \mu\alpha^{(T_2)} \right)}  \bigg\{ \frac{\alpha^{(t)} (1- L \alpha^{(t)})}{2} \left\|\mathbb{E}\left[\boldsymbol{\varepsilon}^{(t)}_g\right]\right\|^{2} \nonumber\\
+&\frac{L (\alpha^{(t)})^{2}}{2} \mathbb{E}\left[\left\|\boldsymbol{\varepsilon}^{(t)}_g\right\|^{2}\right] 
+ \frac{L (\alpha^{(t)} )^{2}}{2} \mathbb{E}\left[\left\|\overline{\mathbf{g}}^{(t)}\right\|^{2}\right] \bigg\}. 
\end{align}

By fixing $\alpha^{(t)} \equiv \alpha$, we have
\begin{align} 
&\mathbb{E}\left[F\left(\mathbf{w}^{(T_2+1)}\right)\right]-F(\mathbf{w}^ *) \nonumber \\
\leq & \left(1- \mu\alpha \right)^{(T_2-T_1)}\left(\mathbb{E}\left[F\left(\mathbf{w}^{(T_1+1)}\right)\right]-F(\mathbf{w}^ *)\right) \nonumber \\
+& \! \! \! \! \sum_{t=T_1+1}^{T_2} \left(1- \mu\alpha \right)^{(T_2-t)}  \bigg\{ \frac{\alpha (1- L \alpha)}{2} \left\|\mathbb{E}\left[\boldsymbol{\varepsilon}^{(t)}_g\right]\right\|^{2} \nonumber\\
+& \frac{L \alpha^{2}}{2} \mathbb{E}\left[\left\|\boldsymbol{\varepsilon}^{(t)}_g\right\|^{2}\right]
+ \frac{L \alpha^{2}}{2} \mathbb{E}\left[\left\|\overline{\mathbf{g}}^{(t)}\right\|^{2}\right] \bigg\}. 
\end{align}
Hence we can obtain the optimality gap in Theorem 1.

\section*{Appendix B: \textsc{Two Kinds of Extensions for Device Scheduling}}
Two kinds of extensions to include the device scheduling are shown as follows.

Case 1: We can simply construct a bicriterion problem that minimizes the weighted sum of the optimality gap and maximizes the cardinality of the currently selected device set. By adding a term $-w^{(t)} |\mathcal{K}^{(t)} |$ to the objective function, where $w^{(t)}>0$ is the parameter to achieve a trade-off between optimality gap and device participation, we can obtain a similar problem as \cite{FL_Multi_IRS} that minimize the optimality gap caused by wireless channel while selecting as more devices as possible for convergence accelerating. 

Case 2: We can mathematically characterize the gradient residual due to the device selection and then formulate the problem. Consider the general case that the $k$-th device has $|\mathcal{D}_k|$ training samples with $\sum_k |\mathcal{D}_k| = |\mathcal{D}|$. Hence, the desired global gradient is given by
\begin{equation} 
	\mathbf{g}^{(t)} = \frac{1}{|\mathcal{D}|} \sum_{k=1}^K |\mathcal{D}_k| \mathbf{g}_{k}^{(t)},
\end{equation}
shown as the weighted average of local gradients with the weights proportional to the size of the corresponding local dataset $ |\mathcal{D}_k|$. Let $\mathcal{K}^{(t)}$ denotes the selected devices set in round $t$, and $\widetilde{\mathcal{K}}^{(t)}$ denotes its complement. Then gradient residual due to the device selection is
\begin{align}
	\bm{\varepsilon}_d^{(t)} \!=\! \frac{1}{\sum_{i \in {\mathcal{K}}^{(t)}} |\mathcal{D}_k|}\sum_{i \in {\mathcal{K}}^{(t)}} |\mathcal{D}_k| \bm{g}_i^{(t)} \! - \! \frac{1}{|\mathcal{D}|} \sum_{k=1}^K |\mathcal{D}_k| \mathbf{g}_{k}^{(t)}.
\end{align}
In that case, the gradient error vector is given by
\begin{align}
    \widetilde{\bm{\varepsilon}_g}^{(t)} = \bm{\varepsilon}_g^{(t)} + \bm{\varepsilon}_d^{(t)},
\end{align}
and we have \cite{SIAM12}
\begin{align}
		\mathbb{E}[\| \widetilde{\bm{\varepsilon}_g}^{(t)} \|^2 ] &\leq  \mathbb{E}[\| {\bm{\varepsilon}_g}^{(t)} \|^2 ] + \mathbb{E}[\|{\bm{\varepsilon}_d}^{(t)} \|^2 ] \nonumber \\
		& \! \leq \! \mathbb{E}[\| {\bm{\varepsilon}_g}^{(t)} \|^2 ] \! + \! \frac{4(|\mathcal{D}| -  \sum_{i \in {\mathcal{K}}^{(t)}} |\mathcal{D}_k|)}{|\mathcal{D}|^2} \gamma^{(t)},
\end{align}
where $\gamma^{(t)}$ is the upper bound of sample-wise gradient. In that case, minimizing the optimality gap consists of two aspects, one is decreasing the number of selected devices to reduce the gradient aggregation error $\bm{\varepsilon}_g^{(t)}$, and the other is increasing the number of selected devices to reduce the gradient residual $\bm{\varepsilon}_d^{(t)}$ due to the device selection.

\section*{Appendix C: \textsc{Proof of Theorem 2}}

Without loss of generality, we adopt the quadratic Lyapunov function $L(\bm{e}(t)) \triangleq \frac{1}{2} \sum_{k \in \mathcal{K}} e_k^2(t)$. Besides the $1$-round Lyapunov drift and the $R$-round Lyapunov drift are represented as $\Delta_1(t)  \triangleq \ L(\bm{e}(t+1)) - L(\bm{e}(t)) $ and $\Delta_R(t)  \triangleq L(\bm{e}(t+R)) - L(\bm{e}(t)) $, respectively. According to \eqref{queue}, we have 
\begin{align}
	&\frac{1}{2} \sum_{k \in \mathcal{K}} e_k^2(t+1) \leq \frac{1}{2}\sum_{k \in \mathcal{K} } \left ( d |b_k^{(t)}|^2 - d{P}_{k}^{\mathrm{avg}} + e_k(t) \right)^2 \nonumber \\
	&\leq  \! C_e \! + \! \frac{1}{2}\sum_{k \in \mathcal{K}} e_k^2(t) \! + \! \sum_{k \in \mathcal{K}} e_k(t)\left (d|b_k^{(t)}|^2 - d{P}_{k}^{\mathrm{avg}} \right),
\end{align}
where $C_e = \frac{1}{2} K E^2_{\max}$ with $ E_{\max} = \max_{k,t} (d|b_k^{(t)}|^2 - d{P}_{k}^{\mathrm{avg}})$. Thus,
\begin{align}
	\Delta_1(t) &\leq C_e + \sum_{k \in \mathcal{K}} e_k(t) \left ( d|b_k^{(t)}|^2 - d{P}_{k}^{\mathrm{avg}} \right) \nonumber \\
	&\leq C_e + \sum_{k \in \mathcal{K}} e_k(t) d\left ( {P}_{k}^{\mathrm{max}} - {P}_{k}^{\mathrm{avg}} \right).
\end{align}

For arbitrary $r$-th period with $V_r$, we have
\begin{align}
	&\Delta_R^\dag(r\rho) + V_r \sum_{t=r\rho+1}^{(r+1)\rho} G_t^\dag \nonumber \\
	\leq & \rho C_e \! + \! \sum_{t=r\rho+1}^{(r+1)\rho}\sum_{k \in \mathcal{K}} e_k(t) \left ( d|b_k^{(t)}|^2 - d{P}_{k}^{\mathrm{avg}}\right) \! + \! V_r \! \! \sum_{t=r\rho+1}^{(r+1)\rho} \! \!G_t^* \nonumber \\
	\leq & \rho C_e + \sum_{t=r\rho+1}^{(r+1)\rho}\sum_{k \in \mathcal{K}} E_{\max}\big((t-1)E_{\max} + e_k(r\rho+1)\big) \nonumber \\
	& + V_r \sum_{t=r\rho+1}^{(r+1)\rho} G_t^*.
\end{align}
Note that $\Delta_R^\dag(r\rho)\geq 0$, thus 
\begin{align}
	 \sum_{t=r\rho+1}^{(r+1)\rho} G_t^\dag &\leq  \sum_{t=r\rho+1}^{(r+1)\rho} G_t^* + \frac{C_r}{V_r},
\end{align}
where $ C_r = \rho C_e + \sum_{t=r\rho+1}^{(r+1)\rho}\sum_{k \in \mathcal{K}} E_{\max}\big((t-1)E_{\max} + e_k(r\rho+1)\big) $. Summing over all the $R$ periods, we can obtain the \eqref{Lyapunov_theorem1} in Theorem 2.


Note that 
\begin{align}
	\Delta_R(t)  \triangleq L(\bm{e}(t+R)) - L(\bm{e}(t)) = \sum_k \frac{1}{2} e^2_k(R+1).
\end{align}
Hence, we have
\begin{align}
	 &\sum_{t=r\rho+1}^{(r+1)\rho} \left (d|b_k^{(t)}|^2 - d{P}_{k}^{\mathrm{avg}}\right) \leq \sum_{t=r\rho+1}^{(r+1)\rho} \big(e_k(t+1) - e_k(t)\big) \nonumber \\
	 & = e_k((r+1)\rho+1) - e_k(r\rho+1) \leq \sqrt{2\Delta_\rho} - e_k(r\rho+1) \nonumber \\ 
	&\leq \sqrt{2\bigg(C_r + V_r \sum_{t=r\rho+1}^{(r+1)\rho} G_t^*\bigg)} - e_k(r\rho+1).
\end{align}
Summing over all the $R$ periods, we can obtain the  \eqref{Lyapunov_theorem2} in Theorem 2.
  
\bibliographystyle{IEEEtran}

\bibliography{refs}

\end{document}